\documentclass[twocolumn,showpacs,preprintnumbers,amsmath,amssymb]{revtex4}

\usepackage{graphicx}
\usepackage{dcolumn}
\usepackage{bm}

\newcommand\beq{\begin{eqnarray}}
\newcommand\eeq{\end{eqnarray}}

\def\qvec{\mbox{\boldmath $q$}_\perp}

\def\kvec{\mbox{\boldmath $k$}_\perp}
\def\k3vec{\mbox{\boldmath $k$}}
\def\bvec{\mbox{\boldmath $b$}}
\def\lvec{\mbox{\boldmath $l$}_\perp}

\def\vepvec{\mbox{\boldmath $\vep$}_\perp}

\def\Delvec{\mbox{\boldmath $\Delta$}_\perp}
\def\0vec{\mbox{\boldmath $0$}_\perp}

\def\vep{\varepsilon}

\def\slash#1{\rlap/{#1}}

\def\pslash{\rlap/{\mkern-1mu p}}
\def\pslash{\slash{\mkern-1mu p}}

\def\eslash{\slash{\mkern-1mu e}}
\def\vepslash{\slash{\mkern-1mu \varepsilon}}

\def\Vslash{\slash{\mkern-1mu {\cal V}}}

\def\nutilde{s}
\def\xp{\xi}

\def\Qbar{\overline{Q}}

\def\es{s}

\input{epsf.sty}
\input{epsfig.sty}

\begin{document}


\title{
Transverse quark motion inside charmonia 
in diffractive photo- and electroproductions
}

\author{A. Hayashigaki$^{1}$ and 
K. Tanaka$^{2}$}
\affiliation{$^{1}$ Theoretische Physik, Universit\"at Regensburg,
D-93053 Regensburg, Germany}
\email{arata.hayashigaki@physik.uni-regensburg.de}
\affiliation{$^{2}$ Department of Physics, Juntendo University, Inba-gun, 
Chiba 270-1695, Japan}
\email{tanakak@sakura.juntendo.ac.jp}

\date{\today}

\begin{abstract}
We reexamine the Fermi motion effects 
on the diffractive photo- and electroproductions 
of heavy vector-mesons, $J/\psi$ and $\psi'$, off a nucleon 
in the leading $\ln (Q^2/\Lambda_{\rm QCD}^{2})$ approximation (LLA) of QCD.
We take into account all the Fermi motion corrections
arising from the relative motion of quarks inside
the charmonium, which is treated as a nonrelativistic bound state
of $c$ and $\bar{c}$.
Our key ingredients are the correct spin structure for the $^{3}S_{1}$ $c\bar{c}$ bound state,
the off-shellness in the $c\bar{c} \rightarrow J/\psi\ (\psi')$ hadronization vertex,
and the modification of 
the gluon's longitudinal momentum fraction 
probed by the process,
due to the relative motion between $c$ and $\bar{c}$.
We demonstrate that these three contributions produce the new Fermi motion effects
in the LLA diffractive amplitude in QCD.
It is found that our new effects moderate the strong suppression of the diffractive $J/\psi$
($\psi'$) production
cross sections, which was reported in the previous works on the Fermi motion effects.
We emphasize the role of the transverse quark motion for the heavy meson production, 
and also discuss the strong helicity dependence of the Fermi motion effects
and its implication in the longitudinal to transverse production ratio, $\sigma_L /\sigma_T$.
\end{abstract}


\maketitle


\section{Introduction}
High-energy diffractive photo- and electroproductions 
of charmonia off a nucleon,
$\gamma^{(*)}+N \rightarrow V+N$ ($V = J/\psi, \psi'$),
are of particular interest,
since they open a new window on
the interface between perturbative 
and nonperturbative 
aspects in QCD \cite{Ryskin,BFGMS,FKS,RRML}.
With perturbative QCD (pQCD),
it has been shown \cite{BFGMS}
that the cross sections of these processes
depend quadratically on the relevant hadronic matrix elements, i.e.,
the wave function (WF) of a charmonium ($J/\psi$ or $\psi'$)
and the gluon distribution inside the target nucleon.
Thus, the diffractive charmonium productions are
expected to be sensitive to those nonperturbative matrix elements,
especially the behavior of the gluon distribution for small Bjorken-$x$.
Recent data from the $ep$ collider HERA 
\cite{H199,ZEUS99,H100,ZEUS97,ZEUS02,H198,H102} 
will indeed 
provide us with quantitative information on them,
while the precision of the data calls for 
QCD analysis 
taking into account also the subleading effects.

\begin{figure}[h]
\vspace*{0cm}
\begin{center}
\hspace*{-3cm}
\psfig{file=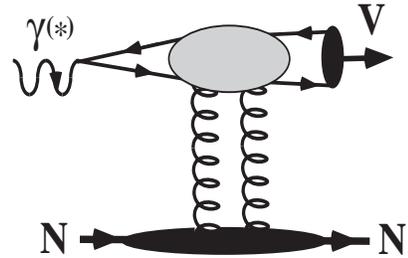,width=9cm,height=3.6cm}
\end{center}
\vspace{0cm}
\caption{Schematic diagram for the
diffractive photo- or electroproduction of a charmonium $V (=J/\psi, \psi')$
off a nucleon $N$, $\gamma^{(*)}+N\rightarrow V+N$.
}
\label{fig:diagram0}
\end{figure}

For high $\gamma^{(*)}$-$N$ center-of-mass energy $W$,
the corresponding diffractive amplitude 
obeys factorization
and, in the leading logarithmic approximation (LLA) of pQCD,
is expressed schematically as (see Fig.~\ref{fig:diagram0})
\begin{eqnarray}
{\cal M}= {\Psi^{V}}^* \otimes 
{\cal A}^{c\bar{c}N} \otimes \Psi^{\gamma} ,
\label{eqn:a}
\end{eqnarray}
with the photon light-cone WF
$\Psi^{\gamma}$
for the $\gamma^{(*)}\rightarrow c\bar{c}$ transition, 
the amplitude
${\cal A}^{c\bar{c}N}$ for the elastic scattering 
of the $c\bar{c}$-pair 
off the nucleon
by exchanging reggeized two gluons,
and 
the vector-meson WF $\Psi^{V}$ 
for the soft hadronization process $c\bar{c}\rightarrow V$ \cite{BFGMS}.
Here, ${\cal A}^{c\bar{c}N}$ is further factorized 
into the $c\bar{c}$-gluon hard scattering amplitude ${\cal A}^{c\bar{c}g}$
and the gluon distribution $G$ inside the nucleon, 
\begin{equation}
{\cal A}^{c\bar{c}N}= 
{\cal A}^{c\bar{c}g} \otimes G .
\label{eqn:a2}
\end{equation}
Physically, the factorization of Eqs.~(\ref{eqn:a}) and (\ref{eqn:a2}) 
follows from the property that 
the typical timescale of the $c\bar{c}$-$N$ scattering at high $W$
is much shorter than the lifetime of the $c\bar{c}$ fluctuation, as well as
of the fluctuation of partons inside the nucleon \cite{BFGMS,RRML}. 
Moreover,
the heavy-quark mass $m_{c}$ ensures
that the $\gamma^{(*)}\rightarrow c\bar{c}$ fluctuation
is sufficiently a short-distance process that takes place 
within a distance of order $1/m_{c}$ or less.
This enables us to treat the photon WF $\Psi^{\gamma}$,
as well as the scattering amplitude ${\cal A}^{c\bar{c}g}$
between the small-size $c\bar{c}$-color-dipole and the gluon,
by perturbation theory even for the photoproduction with the 
photon's virtuality $Q^{2}=0$.
%
%

Eq.~(\ref{eqn:a}) involves  
the convolution with the $c\bar{c}$-bound-state WF
$\Psi^{V}$.
Here, the large mass of
$m_{c}$ also allows us to organize the amplitude ${\cal M}$
into a hierarchy in terms of the dependence on the average velocity $v$ of the heavy quark 
in the charmonium rest frame.
The leading term independent of $v$ corresponds to the static or nonrelativistic limit,
where $c$ and $\bar{c}$ have no relative motion inside the charmonium,
while the subleading terms in powers of $v$ 
express the Fermi motion corrections and relativistic corrections.
For the charmonium, quark potential model calculations \cite{QR} 
as well as lattice QCD simulations \cite{MILC}
indicate that 
$v$ is reasonably small, $v^{2}\simeq 0.2\sim 0.3$.
Thus, the subleading terms are not expected to produce
so large corrections, 
though they could be physically important.

So far, there exist several investigations for the Fermi motion corrections,
but significant discrepancies are observed among the previous results: 
Frankfurt-Koepf-Strikman (FKS) \cite{FKS} predicted 
quite strong suppression of the cross section 
for the diffractive photo- (electro-)production of $J/\psi$
by a factor $\lesssim 0.1$ ($\lesssim 0.3$) 
due to the Fermi motion effects.
On the other hand, the results of Ryskin-Roberts-Martin-Levin (RRML) \cite{RRML} 
gave a significantly weaker suppression factor of $0.4 \sim 0.6$
for the $J/\psi$ photoproduction.
Furthermore, the author of Ref.~\cite{PH} estimated
an even weaker suppression factor of 0.9 or more
for the $J/\psi$ electroproduction.
The treatment of these three works is different in several points; 
in particular, the difference in the detailed shape of the meson WF $\Psi^{V}$ 
appears to be an important source of the discrepancies,
as claimed by the authors of Ref.~\cite{FKS}.
In our opinion, however, any of those works is not 
satisfactory as the calculations of the Fermi motion effects,
because certain effects due to the relative motion of the quarks
are not taken into account.

The aim of this paper is to 
reexamine the Fermi motion corrections on 
the diffractive photo- and electroproductions
of charmonia.
Similarly to the previous works \cite{FKS,RRML},
our basic formulation relies on the QCD factorization approach in the LLA of pQCD.
In this framework, we re-derive the formulae (\ref{eqn:a}), (\ref{eqn:a2}) explicitly,
paying attention to the contributions that are related to the relative motion 
of the quarks and thus 
should be regarded as 
a part of the Fermi motion effects.
We find three types of such additional contributions, which 
have not been considered properly in a single treatment so far.

First of all, from the viewpoint of the systematic expansion 
of the total amplitude (\ref{eqn:a}) in terms of the velocity $v$,
the Fermi motion corrections through the relative order $v^{2}$
should be computed with the Bethe-Salpeter $c\bar{c}$ WF $\Psi^{V}$, which
is defined as the matrix element of a bilocal operator composed 
of the two-component Pauli spinor fields for $c$ and $\bar{c}$
in a nonrelativistic Schr\"{o}dinger field theory,
replacing the original relativistic field theory. 
Namely, when the WF $\Psi^{V}$ is expressed 
in the usual four-component Dirac representation,
it is subject to a constraint
${\cal R}\Psi^{V} = \Psi^{V}$ with the projector ${\cal R}$ onto the ``large components''
corresponding to the Pauli spinor.  
Physically, this implies that 
the WF $\Psi^{V}$ describes a pure $^{3}S_{1}$ $c\bar{c}$-state.
RRML did not consider the role of such projector ${\cal R}$ at all \cite{RRML}.
In FKS \cite{FKS}, an improper projector ${\cal R}$ was used, 
which was constructed in an oversimplified fashion.
We employ the projector ${\cal R}$, which obeys spin-symmetry
in the Schr\"{o}dinger field theory and coincides with that appeared in Ref. \cite{PH}.
We will show that the spin structure of $\Psi^{V}$, 
which is imposed by our ${\cal R}$, modifies 
the amplitude (\ref{eqn:a}) at the order $v^{2}$ in comparison with RRML or FKS.

Following the previous works \cite{BFGMS,FKS,RRML},
when combining the factors of Eq.~(\ref{eqn:a}),
we express the meson WF $\Psi^{V}$, as well as the other parts 
${\cal A}^{c\bar{c}N}$ and $\Psi^{\gamma}$,
in a helicity representation.
This corresponds to a familiar representation of the QCD factorization formulae
in the Lepage-Brodsky approach for hard exclusive processes \cite{LB}. 
The helicity representation of ${\Psi^{V}}^{*}$ in Eq.~(\ref{eqn:a}) 
is given by the matrix element
$\bar{v}_{\lambda'}{\Psi^{V}}^{*}u_{\lambda}$,
in terms of helicity Dirac spinors $u_{\lambda}$ and $v_{\lambda'}$ for quark and antiquark.
This spinor matrix element representing the effective ``$c\bar{c}\rightarrow V$ vertex'' 
may be off the energy shell:
the total energy carried by the two spinors $\bar{v}_{\lambda'}$ and $u_{\lambda}$
is generally different from the energy of the vector meson.
In the previous works \cite{FKS,RRML}, however, 
the on-shellness was assumed when evaluating the matrix element
for the effective $c\bar{c}\rightarrow V$ vertex.
Apparently, this assumption corresponds to neglecting the binding energy of the charmonium,
so that it could give the $O(v^{2})$ error in the final result.
Thus, the second of the additional Fermi motion contribution is generated by 
the off-shellness of the $c\bar{c}\rightarrow V$ vertex.

Third, the Feynman diagrams relevant for the LLA of the factorization formula 
 (\ref{eqn:a2}) 
tell us that the gluon's momentum fraction inside the nucleon, 
which is probed by the diffractive production,
depends on the longitudinal as well as transverse relative motion of $c$ and $\bar{c}$ 
through the momentum exchange at the quark-gluon coupling.
Namely,
the Bjorken-$x$ of the distribution $G$ of Eq.~(\ref{eqn:a2}) depends 
on the motion of the quarks, which produces the $O(v^{2})$
effects to the amplitude (\ref{eqn:a}),
though the previous works \cite{FKS,RRML,PH} 
seem to have simply neglected this contribution.

As we will demonstrate, the modification of the amplitude (\ref{eqn:a}) due to
the above three types of contributions is sensitive to
the transverse relative motion of the quarks inside charmonia; i.e., the transverse quark motion is no less
important than the longitudinal one for the production of nonrelativistic bound-state 
with $Q^{2} \lesssim m_{c}^{2}$.
This is in contrast with the light-cone dominated processes such as
the hard diffractive productions of the $\rho$ meson ($Q^2 \gg \Lambda_{\rm QCD}^{2}$),
where the longitudinal quark motion plays a dominant role compared with 
the transverse one, and the effects of the transverse quark motion 
give only small corrections of higher twist $\sim \Lambda_{\rm QCD}^{2}/Q^{2}$ \cite{BFGMS}.

By including those three types of additional contributions,
we will give quantitative estimates of the Fermi motion effects.
We employ the ``shape'' of the meson WF $\Psi^{V}$, 
which is obtained by solving the Schr\"{o}dinger
equation with a realistic potential between the quark and the antiquark.
We compute the cross sections for the diffractive photo- and electroproductions
of $J/\psi$,
and find that the Fermi motion effects give moderate suppression.
The net effect of our Fermi motion contributions is, roughly speaking, 
similar to that of RRML \cite{RRML},
but our results show some new features in comparison with those of RRML as well as of Refs. \cite{FKS,PH}. 
To demonstrate them, we calculate the ratio of the cross sections with longitudinally and
transversely polarized photons, $\sigma_L/\sigma_T$.
Because the contribution of the gluon distribution $G$ proves to mostly cancel in this ratio,
this quantity is sensitive to the first two 
of the above-mentioned additional Fermi motion effects,
which are strongly helicity-dependent.
Another interesting quantity is the ratio of $\psi'$ to $J/\psi$ cross sections, by
extending our calculation to the case of a radial excitation $\psi'$.
We observe that the Fermi motion effects produce characteristic behavior
for the ratio, reflecting the different shape of the WFs of $J/\psi$ and $\psi'$.

The paper is organized as follows:
in Sec.~II, we derive the formulae (\ref{eqn:a}) and (\ref{eqn:a2}) in the LLA
with our additional Fermi motion effects. 
After recalling the basic steps to derive the factorization formula 
for the diffractive production, 
the three types of additional Fermi motion effects are explained in detail in Secs.~II~A-C.
The final form of our production cross sections is given in Sec.~II~D.
Sec.~III contains our numerical results. 
In Sec.~III~A, we present $\sigma_{L}$, $\sigma_{T}$ and $\sigma_{L}/\sigma_{T}$ 
for the $J/\psi$ productions, demonstrating roles of each of our three corrections.
We show the total cross sections for the $J/\psi$ photo- and electroproductions
in Sec.~III~B, making a comparison with the HERA data,
and clarify the connection with the previous works.
Similar program is carried out for the $\psi'$ productions in Sec.~III~C, 
and also the implication of our results in recent debate \cite{HP,SHIAH,HIKT} 
on the $\psi'$ to $J/\psi$ ratio is discussed.
In Sec.~IV, a summary and discussion are presented.

\section{Formulation}
\subsection{The forward diffractive amplitude in QCD}

\begin{figure}[h]
\vspace*{0cm}
\begin{center}
\hspace*{0cm}
\psfig{file=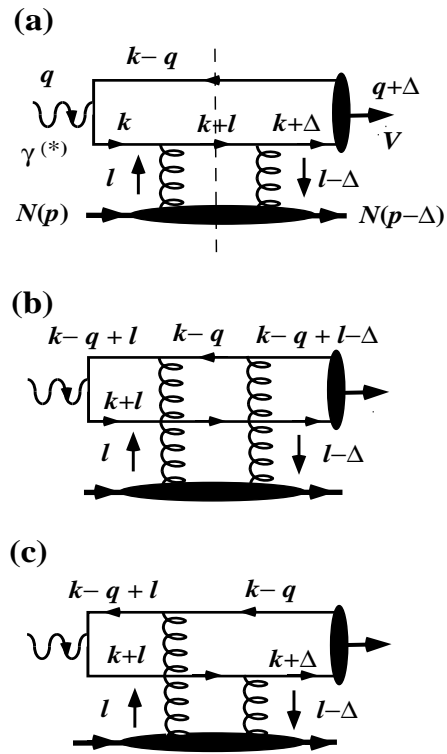,width=7cm,height=10cm}
\end{center}
\vspace{0cm}
\caption{Feynman diagrams for the
diffractive photo- or electroproduction of $V=J/\psi, \psi'$.
The dashed line means the discontinuity of the amplitude.
}
\label{fig:diagram}
\end{figure}

Let us first describe a basic formulation
to evaluate the diffractive photo- or electroproduction of the heavy quarkonium.    
We consider the forward diffractive production of the heavy vector-meson, 
$\gamma^{(*)} (q)+ N (p) \to V (q+\Delta) + N (p- \Delta)$, where 
$q$ and $p$ denote the four momenta of the photon and the incoming nucleon, respectively, and 
$\Delta$ the momentum transfer in the $t$-channel.   
The total center-of-mass energy of the $\gamma^{(*)}$-$N$ system,  
$W = \sqrt{(p + q)^2}$, is assumed to be much larger than the photon's virtuality
and the heavy-quark mass,
\begin{equation}
W^{2} \gg Q^{2}, \;\;\;\;\;\;\;\;\;\;\;\; W^{2}\gg m_{c}^{2},
\label{eq:he}
\end{equation}
with $Q^2 = -q^2$.
We also suppose $m_{c}^{2} \gg -t$ with $t\equiv \Delta^2$, 
and $m_{c}^{2} \gg \Lambda_{\rm QCD}^{2}$.
Under such kinematical condition, the relevant diagrams 
for the scattering amplitude in the LLA are shown in Fig.~\ref{fig:diagram}.
%
The heavy-quark mass $m_{c}$ allows us to calculate the creation of the $c\bar{c}$ pair, 
$\gamma^{(*)} \rightarrow c\bar{c}$,
as well as its time development by perturbation theory
before the nonperturbative effects to form the quarkonium state $V$ set in.
As is well-known, the corresponding amplitude is 
dominated by its imaginary part, which we   
shall calculate explicitly;
the contributions from the real part can be incorporated perturbatively.
For simplicity, we consider
the forward limit of the diffractive process, $t \to 0$, in detail.

To proceed, we introduce two
null vectors $q'$ and $p'$ with ${q'}^{2}={p'}^{2}=0$ and $q'\cdot p'\equiv \es/2$ as
\begin{eqnarray*}
q &=& q'-\frac{Q^2}{\nutilde}p',
\\
p &=& \frac{M_N^2}{\nutilde}q'+p',
\end{eqnarray*}
where $M_N$ is the nucleon mass, $p^2 = M_N^{2}$, 
and $\es \cong W^2-M_N^2+Q^2$. 
In the forward limit $t=0$, we also have
\begin{displaymath}
\Delta = \frac{Q^2 + M_{V}^{2}}{\es}p'
\end{displaymath}
with $M_V$ the mass of the heavy vector-meson, 
so that $(p-\Delta)^2 = M_N^{2}$ and $(q+\Delta)^2 = M_V^{2}$. 
In Fig.~\ref{fig:diagram}, the momenta $k$ for the internal quark lines and $l$ 
for the exchanged gluons have the Sudakov decomposition with respect to $q'$ and $p'$:
\begin{eqnarray}
k &=& \alpha q'+\beta p' + k_\perp ,
\label{eq:k}
\\
l &=& \kappa q'+ \xp p' + l_\perp .
\label{eq:l}
\end{eqnarray}
Here, $\alpha$ expresses the fraction of the photon momentum carried by 
the charm quark along the ``$q'$ direction'', while $\xp$ the fraction of the 
nucleon momentum carried by the gluon along the ``$p'$ direction''.
We choose a Lorentz frame as $q^{+}= {q'}^{+}$ and $p^{-}= {p'}^{-}$, so that
$k= (\alpha q^{+}, \beta p^{-}, \kvec)$, $k_\perp^2=-\kvec^2$, and similarly for $l$.

In the calculation of the discontinuity of the relevant diagrams,
the on-shell conditions for the 
quark lines, which cross the ``cut'' as shown in Fig.~\ref{fig:diagram}(a),
determine $\beta$ and $\xp$ as
\begin{eqnarray}
\beta &=& - \frac{1}{\es(1-\alpha)}\left[(1-\alpha) Q^2 + \kvec^2 + m_{c}^2 \right],
\label{eq:beta} \\
\xp &=& \frac{1}{\es} \left[ Q^2 
+\frac{\kvec^2 + m_{c}^2}{1-\alpha}  + \frac{(\kvec + \lvec)^2 + m_{c}^2}{\alpha}\right],
\label{eq:eks}
\end{eqnarray} 
while the integrations over $\alpha$, $\kvec$, $\kappa$, and $\lvec$
has to be performed.
Now, formally, the factorization formula follows from the observation that
the dominant integration region in the $\kappa$-variable
is the one of $\kappa = O(\lvec^2/\es)$ for fixed $\lvec$
in the high-energy limit (\ref{eq:he}),
and, in this regime $l^{2} \simeq -\lvec^{2} \ll \es$,
the exchanged soft gluons couple to the quarks 
with an eikonal vertex $g \pslash'$ corresponding to a single gluon polarization.
The use of the eikonal vertex immediately leads to the factorization of 
the photon WF $\Psi^{\gamma}$ and the vector-meson WF $\Psi^{V}$ 
from ${\rm Im} {\cal M}$ (see Eq.~(\ref{eqn:a})). 
Also, the $\kappa$-integration reduces to that for the nucleon matrix element 
by Taylor expanding explicit $\kappa$-dependence of the other parts in the integrand,
and this accomplishes the factorization of the gluon distribution along with 
appropriate decoupling of the color indices;
this is so-called ``$k_{\perp}$-factorization'' \cite{CCH},
because the $\lvec$-dependence in the integrand has to be retained in the present regime.
As a result, the unintegrated gluon density distribution 
$f(\xi, \lvec^{2})$, depending on the transverse momentum $\lvec$ 
as well as the longitudinal momentum fraction $\xi$,
describes the nonperturbative part due to the nucleon matrix element \cite{RRML}.
(Actually, the unintegrated gluon density $f(\xi, \lvec^{2})$ for $l^{2} \simeq -\lvec^{2}$
involves all multiple gluon exchanges, which give the ``ladder structure'' described 
by the BFKL evolution, but in the LLA we can eventually eliminate $f$
in terms of the conventional gluon distribution $G$ as in  Eq.~(\ref{eqn:a2}).)
For example, we obtain as the discontinuity of Fig.~\ref{fig:diagram}(a),
\begin{eqnarray}
\lefteqn{
\frac{\pi^2 \alpha_s}{\sqrt{2N_c}\ \nutilde}
\int_0^1 \frac{d\alpha}{\alpha^2} \int\frac{d^{2}k_{\perp} d^{2}l_{\perp}}{(2\pi)^4}
\frac{
f(\xi,\lvec^2)}{\lvec^4}}
\nonumber\\
&\times& \sum_{\lambda\lambda'\lambda''\lambda'''}
{\Psi_{\lambda''' \lambda }^{V(\zeta')}}^{*}(\alpha,\kvec) 
\left[\overline{u}_{\lambda'''}(k)\pslash' u_{\lambda''}(k+l)\right]\ 
\nonumber\\
&\times& \left[\overline{u}_{\lambda''}(k+l)\pslash' u_{\lambda'}(k)\right]
\Psi_{\lambda'\lambda}^{\gamma (\zeta)}(\alpha,\kvec),
\label{eqn:defM}
\end{eqnarray}
up to the terms suppressed for high $\es$.
Here, $\alpha_{s}=g^2/(4\pi)$, 
$N_{c}$ is the number of colors, and 
$\zeta$ and $\zeta'$ denote the helicities of the photon and the vector meson, respectively.
$u_{\lambda'}(k)$ is the on-shell spinor for the quark with the helicity $\lambda'$.

The light-cone WF $\Psi^{\gamma(\zeta)}_{\lambda' \lambda}(\alpha,\kvec)$ 
describes the probability amplitude to find the photon of helicity $\zeta$ 
in a state with the minimal number of constituents $c$ and $\bar{c}$,
which carry the longitudinal momentum fractions $\alpha$ and $1-\alpha$,
the transverse momenta $\kvec$ and $-\kvec$,
and the helicity $\lambda'$ and $\lambda$, respectively. 
It is given as the lowest order amplitude for the 
photon dissociation to the $c \bar c$-pair (see Fig.~\ref{fig:diagram}),
and the corresponding formula is well-known \cite{BFGMS}:
\begin{eqnarray}
\Psi^{\gamma (\zeta)}_{\lambda' \lambda} (\alpha,\kvec) = - ee_{c}
\frac{\sqrt{\alpha(1-\alpha)} \bar{u}_{\lambda'}(k) \vepslash^{(\zeta)}
v_{\lambda}(q-k)}
{\alpha (1-\alpha) Q^2 + \kvec^2 +m_c^2},
\label{eq:Psig}
\end{eqnarray}
where 
$v_\lambda(q-k)$ is the on-shell helicity spinor for the antiquark.
$e$ is the charge of the proton and $e_c=2/3$ gives the charge of the $c$-quark.  
The polarization vectors $\vep^{(\zeta)}$ of the photon are
defined as
\begin{eqnarray}
\vep^{(0)} &=& \frac{1}{Q}\left(q'+\frac{Q^2}{\nutilde} p'\right),
\label{eq:phe1}\\
\vep^{(\pm 1)} &=&\vep_{\perp}^{(\pm 1)}=
\left(0,\ 0,\ \frac{1}{\sqrt{2}},\  \pm \frac{i}{\sqrt{2}}\right),
\label{eq:phe2}
\end{eqnarray}
for the longitudinal ($\zeta =0$) and transverse ($\zeta = \pm 1$) polarizations,
respectively.
Calculating the spinor matrix elements in Eq.~(\ref{eq:Psig})
with a helicity basis by Lepage and Brodsky \cite{LB}, one finds
for the longitudinal photon,
\begin{eqnarray}
\lefteqn{\Psi_{\lambda'\lambda}^{\gamma(0)}(\alpha,\kvec)}
\nonumber\\
&=&-ee_c 
\left[\frac{2\alpha(1-\alpha)Q}{\alpha(1-\alpha)Q^2+\kvec^2+m_c^2}
-\frac{1}{Q}\right] \delta_{\lambda',-\lambda},
\label{eqn:photon-wL}
\end{eqnarray}
and, for the transverse photon, 
\begin{eqnarray}
\lefteqn{\Psi_{\lambda'\lambda}^{\gamma(\tau)}(\alpha,\kvec)}
\nonumber\\
&&=
\left\{ {\begin{array}{lr}
 -\frac{\displaystyle{\sqrt{2}\ e e_c \  m_c \ \lambda \delta_{\lambda, \tau}}}
{\displaystyle{\kvec^{2}+m_c^2+\alpha(1-\alpha)Q^2}}
 &\;\; \;(\lambda'=\lambda)  \\
\\
 -\frac{\displaystyle{2 e e_{c}\ (\vepvec^{(\tau)}\cdot\kvec) (\alpha-\delta_{\lambda,\tau})}}
{\displaystyle{\kvec^{2}+m_c^2+\alpha(1-\alpha)Q^2}} 
&\; \; \;\;(\lambda'=-\lambda) \\
\end{array}} \right.
\label{eqn:photon-wT}
\end{eqnarray}
with $\tau = \pm 1$ and $\vepvec^{(\tau)}=\frac{1}{\sqrt{2}}(1,\tau i)$. 
Note that the second term in Eq.~(\ref{eqn:photon-wL}) 
is independent of both $\alpha$ and $\kvec$;
as is well-known, this term can be omitted due to crossing symmetry,
since the contribution from this term is cancelled by the corresponding 
contribution from the diagram (c) 
of Fig.~\ref{fig:diagram} (see Eqs.~(\ref{eq:amp}), (\ref{eqn:A}) below).

In Eq.~(\ref{eqn:defM}),
the formation process of the vector meson $V$ from the outgoing $c\bar{c}$-pair
is factorized as the vector-meson light-cone WF 
${\Psi_{\lambda''' \lambda }^{V(\zeta')}}^{*}(\alpha,\kvec)$.
Its physical interpretation is similar to the photon WF 
$\Psi^{\gamma (\zeta)}_{\lambda' \lambda} (\alpha,\kvec)$,
but ${\Psi_{\lambda''' \lambda }^{V(\zeta')}}^{*}(\alpha,\kvec)$ is more sophisticated
because it contains nonperturbative dynamics between $c$ and $\bar{c}$.
Its explicit formula will be discussed in Sec.~II~B.

We note that the eikonal vertex $\pslash'$ does not change the helicity of quark 
or antiquark for the Lepage-Brodsky spinors:
\begin{eqnarray}
\bar{u}_{\lambda_{1}}(k_{1})\pslash' u_{\lambda_{2}}(k_{2})
&=&
\bar{v}_{\lambda_{1}}(k_{1})\pslash' v_{\lambda_{2}}(k_{2})
\nonumber\\
&=& \frac{\es}{q^{+}}\sqrt{k_{1}^{+}k_{2}^{+}}\ \delta_{\lambda_{1},\lambda_{2}}.
\label{eq:eik}
\end{eqnarray}
Substituting this 
into Eq.~(\ref{eqn:defM}),
we immediately get the $s$-channel helicity conservation, i.e.,
Eq.~(\ref{eqn:defM}) is nonzero only for $\zeta = \zeta'$.

The contributions from the diagrams (b) and (c) of Fig.~\ref{fig:diagram}
can be computed in a similar way to the diagram (a). The results are expressed 
with the same light-cone WFs $\Psi^{\gamma (\zeta)}_{\lambda' \lambda}$,
${\Psi_{\lambda''' \lambda }^{V(\zeta')}}^{*}$
and gluon distribution $f$ as in Eq.~(\ref{eqn:defM}),
and satisfy the $s$-channel helicity conservation.
Combining those contributions with Eq.~(\ref{eqn:defM}),
we get the discontinuity of the total diffractive amplitude for the helicity
$\zeta$ of the incoming photon as 
\begin{eqnarray}
{\rm Im}{\cal M}^{(\zeta)}
&=&\frac{3\es}{2\sqrt{2N_c}} \sum_{\lambda'\lambda}\int \frac{d\alpha d^2 k_{\perp}}{16\pi^3}
\nonumber\\
&&\times{\Psi^{V(\zeta)}_{\lambda' \lambda}}^{*}(\alpha,\kvec)
\hat{\sigma}^{c\bar{c}N}\Psi^{\gamma(\zeta)}_{\lambda' \lambda}(\alpha,\kvec) ,
\label{eq:amp}
\end{eqnarray}
where $\hat{\sigma}^{c\bar{c}N}$
expresses the operator acting on 
$\Psi^{\gamma(\zeta)}_{\lambda' \lambda}(\alpha,\kvec)$:
\begin{equation}
\hat{\sigma}^{c\bar{c}N}= \frac{2 \pi \alpha_s}{3}
\int d^{2}l_{\perp}\ 
\frac{f (\xi, \lvec^2)}{\lvec^4}
\left( 1+ 1 - 2e^{i\lvec \cdot \hat{\bvec}}
\right) ,
\label{eqn:A}
\end{equation}
with $\hat{\bvec} \equiv i \partial /\partial \kvec$. 
Three terms in the integrand correspond
to the diagrams (a), (b), and (c) of Fig.~\ref{fig:diagram}, respectively.
Note that Eq.~(\ref{eq:amp})
obeys the standard representation of the factorization formula 
in the Lepage-Brodsky approach for the exclusive processes \cite{LB}.

In Eqs.~(\ref{eq:amp}), (\ref{eqn:A}), the $\lvec$ integral converges 
for $|\lvec| \rightarrow \infty$.
Moreover, in the integrand, we have the contributions behaving like
\begin{equation} 
\frac{1}{\kvec^2+\Qbar^2}
-\frac{1}{(\kvec-\lvec)^2+\Qbar^2}  ,
\label{eq:com}
\end{equation}
with $\Qbar^2=m_c^2+\alpha(1-\alpha)Q^2$,
from the denominator of $\Psi^{\gamma(\zeta)}_{\lambda' \lambda}(\alpha,\kvec)$ 
of Eq.~(\ref{eq:Psig}).
Then, the important domain in the $\lvec$-integral is
$\lvec^2 \lesssim \Qbar^2$, and, in particular, a logarithmic contribution 
$\sim \ln (\Qbar^2/\Lambda_{\rm QCD}^{2})$ comes from $\lvec^{2} \ll \Qbar^2$.
Here, up to non-logarithmic corrections,
Eq.~(\ref{eqn:A}) can be simplified by the replacement
$1 -  e^{i\lvec \cdot \hat{\bvec}} \rightarrow
\lvec^{2}\ \hat{\bvec}^2 /4$, retaining only the leading nonzero term in 
the power series of $\lvec\cdot \hat{\bvec}$,
which corresponds to the ``color-dipole picture'' \cite{FKS,SHIAH}.
In the same accuracy, we neglect the $\lvec$ dependence of
$\xi$ in $f(\xi, \lvec^2)$ (see Eq.~(\ref{eq:eks})),
i.e., replace $\xi$ by
\begin{equation}
\tilde{x} = \frac{1}{\es} \left( Q^2 
+\frac{\kvec^2 + m_{c}^2}{\alpha(1-\alpha)} \right) .
\label{eq:tilx}
\end{equation}
Now, using the relation 
\begin{equation}
\tilde{x}G(\tilde{x},\Qbar^2)=\int^{\Qbar^2} d \lvec^2 \frac{f(\tilde{x},\lvec^2)}{\lvec^2}
\label{eqn:unintegratedg}
\end{equation}
with the conventional gluon distribution $G$,
we get, in the leading logarithmic accuracy,
\begin{equation}
\hat{\sigma}^{c\bar{c}N}= \frac{\pi^2}{3}\alpha_s(Q_{\rm eff}^2)
\tilde{x}G(\tilde{x},Q_{\rm eff}^2) \hat{\bvec}^{2} ,
\label{eqn:A2}
\end{equation}
with $Q_{\rm eff}^2 = (Q^2 + M^{2}_{V})/4$.
This operator corresponds to the color-dipole cross section,
representing the high-energy interaction of a small-size
quark-antiquark configuration with a target.
The results (\ref{eq:amp}) and (\ref{eqn:A2}) demonstrate the factorization
(\ref{eqn:a}) and (\ref{eqn:a2}); note that, to derive those results,
we have not employed any other assumption than the high-energy limit and the LLA.
Eqs.~(\ref{eq:amp}), (\ref{eqn:A2}) coincide with the corresponding results of 
Refs. \cite{BFGMS,FKS,RRML}, except that $\tilde{x}$ of Eq.~(\ref{eq:tilx})
appears
instead of the usual $x=\left( Q^2 + M_{V}^{2} \right)/\es$,
and that we shall use a more precise form for the heavy vector-meson WF 
${\Psi^{V(\zeta)}_{\lambda' \lambda}}(\alpha,\kvec)$.
In the following subsections, we will explain 
that those differences produce the ``new'' Fermi motion corrections
of order $v^{2}$ that were mentioned in Sec.~I, 
when expanded in powers of the heavy-quark velocity
$v$ inside the vector meson.

\subsection{Spin structure of heavy-meson WF}
The light-cone WF of the vector mesons 
is defined as the Fourier transform of the Bethe-Salpeter $c\bar{c}$
WF at equal light-cone time $x^{+}=0$.
It is a nonperturbative object, which is not fully understood at present.  
In the same spirit as in Refs. \cite{FKS,RRML}, 
we shall construct the shape of the heavy-meson WF
by solving the Schr\"{o}dinger equation
with a realistic potential between quark and antiquark.
As we will demonstrate below, 
this indeed gives a convenient framework to evaluate the Fermi motion effects
of order $v^{2}$ due to the relative motion of the quarks
inside the nonrelativistic bound state.

A field-theoretic definition of the corresponding WF is provided by
the heavy-quark limit of the original Bethe-Salpeter WF, 
i.e., by the Bethe-Salpeter $c\bar{c}$ WF in terms of the  
two-component Pauli-spinor fields in a nonrelativistic Schr\"{o}dinger field theory.
Spin symmetry tells us that the $S$-wave $c\bar{c}$ states form the quartet 
consisting of the spin singlet and the spin triplet, 
and the spin triplet states describe the vector mesons like $J /\psi, \psi'$. 
As is well-known \cite{BJ}, the corresponding coupling of the heavy-quark spins is represented
by the matrices,
when expressed in the original four-component Dirac representation,
\begin{equation}
{\cal R} \gamma_{5}, \;\;\;\;\;\;\;\;\;\;\;\; 
{\cal R} \eslash^{(\zeta)} ,
\label{eq:spinV}
\end{equation}
for the spin singlet and triplet states, respectively, with
the projector ${\cal R}$ onto the ``large components''
corresponding to the Pauli spinor: 
\begin{eqnarray}
{\cal R} \equiv 
\frac{1+\Vslash}{2}.
\label{eqn:projection}
\end{eqnarray}
Here, ${\cal V}$ denotes the four velocity of the relevant mesons,
and $e^{(\zeta)}$ denotes the 
polarization vector of the vector meson.
Thanks to spin symmetry, the matrices (\ref{eq:spinV}) 
completely determine the spin structure of the 
light-cone WFs for the pseudoscalar and vector mesons in the heavy-quark limit.
In a recent work \cite{AST},
the diffractive $\eta_{c}$ and $\eta_b$
productions induced by neutrino beam 
have been discussed using the first of the spin WFs (\ref{eq:spinV}).
Now, the second one with ${\cal V}=(q+\Delta)/M_{V}$ gives our light-cone WF 
for the vector meson.
In the helicity $(\lambda', \lambda)$-representation, it reads
\begin{eqnarray}
{\Psi^{V(\zeta)}_{\lambda' \lambda}}^{*}(\alpha,\mbox{\boldmath $k$}_\perp)
&=&\frac{\bar{v}_{\lambda}(q-k)}{\sqrt{1-\alpha}}\gamma^{\mu}
{e_{\mu}^{(\zeta)}}^{*} {\cal R}
\frac{u_{\lambda'}(k)}{\sqrt{\alpha}}\frac{\phi^*(\alpha,\kvec)}{M_V}
\nonumber\\
\label{eqn:wfV}
\end{eqnarray}
with the Lepage-Brodsky spinors $u_{\lambda'}(k)$ and $v_{\lambda}(q-k)$.
The polarization vectors ${e_{\mu}^{(\zeta)}}$ are given as 
\begin{eqnarray}
e^{(0)} &=& \frac{1}{M_{V}}\left(q' - \frac{M_{V}^{2}}{\es}p'\right),
\label{eq:el}\\
e^{(\pm 1)} &=&e_{\perp}^{(\pm 1)}=
\left(0,\ 0,\ \frac{1}{\sqrt{2}},\  \pm \frac{i}{\sqrt{2}}\right),
\label{eq:et}
\end{eqnarray}
for the longitudinal ($\zeta = 0$) and transverse ($\zeta = \pm 1$) polarizations,
respectively, so that  
$e^{(\zeta)}\cdot {\cal V}=0$ and 
${e^{(\zeta)}}^{*}\cdot e^{(\zeta')} = - \delta_{\zeta, \zeta'}$.
The scalar function $\phi^*(\alpha,\kvec)$ represents the nonperturbative part provided
by a $S$-wave solution of the Schr\"{o}dinger equation,
whose detailed form will be specified later.  
We note that the spin structure of Eq.~(\ref{eqn:wfV}) reduces to that of the light-cone WF 
for the light vector-meson of Ref. \cite{BFGMS}
by the replacement ${\cal R} \rightarrow 1$.

It is straightforward to compute the spinor matrix elements of Eq.~(\ref{eqn:wfV}),
which give the effective
``$c\bar{c} \rightarrow V$ vertex'':
\begin{equation}
\Gamma^{(\zeta)*}_{\lambda' \lambda}(\alpha,\mbox{\boldmath $k$}_\perp)
\equiv
\frac{1}{M_V}
\frac{\bar{v}_{\lambda}(q-k)}{\sqrt{1-\alpha}}\gamma^{\mu}
{e_{\mu}^{(\zeta)}}^{*} {\cal R}
\frac{u_{\lambda'}(k)}{\sqrt{\alpha}}. 
\label{eqn:GamV}
\end{equation}
We get, for the longitudinal polarization,
\begin{eqnarray}
\Gamma^{(0)*}_{\lambda \lambda}
&=&
\frac{\lambda \left(2\alpha-1\right)(\vepvec^{(\lambda)}\cdot\kvec)}
{\sqrt{2}\ \alpha(1-\alpha)M_V},
\label{eqn:L1}\\
\Gamma^{(0)*}_{-\lambda \lambda}
&=&
-\frac{1}{2}-\frac{\kvec^{2}+m_c (m_{c} + M_{V})}
{2\alpha(1-\alpha)M_V^2},
\label{eqn:L2}
\end{eqnarray}
where $\vepvec^{(\lambda)}=\frac{1}{\sqrt{2}}(1,\lambda i)$,
and, for the transverse polarizations ($\tau = \pm1$),
\begin{eqnarray}
\Gamma^{(\tau)*}_{\lambda \lambda}
&=&
\left[\delta_{\tau,\lambda}
\left\{(M_V+m_c)m_c+\alpha(1-\alpha)M_V^2\right\}
\right.
\nonumber \\
&&\!\!\!\!\!\!\!\!
+
\left.2\delta_{\tau,-\lambda}
(\vepvec^{(\tau) *}\cdot \kvec )^2
\right]\frac{\lambda}{\sqrt{2}\ \alpha(1-\alpha)M_V^2},
\label{eqn:T1}\\
\Gamma^{(\tau)*}_{-\lambda \lambda}
&=&
\left[(\alpha-\delta_{\tau,\lambda})(M_V+2m_c)\right.
\nonumber\\
&+& \left.(1-2\alpha)m_c\right]
\frac{(\vepvec^{(\tau)*}\cdot \kvec)}
{\alpha(1-\alpha)M_V^2}.
\label{eqn:T2}
\end{eqnarray}
Using these results, the discontinuity of
the total diffractive amplitude (\ref{eq:amp}) can be expressed in the LLA
as
\begin{equation}
{\rm Im}{\cal M}^{(\zeta)}
= \int_{0}^{1} d\alpha \int d^2 k_{\perp}
\phi^{*}(\alpha,\kvec)\Omega^{(\zeta)}(\alpha,\kvec),
\label{eq:ampnew}
\end{equation}
where $\Omega^{(\zeta)}(\alpha,\kvec)$ represents the ``effective 
$\gamma N V$ vertex'', whose ($\alpha, \kvec)$-dependence is completely determined by
the analytic formula:
\begin{eqnarray}
\lefteqn{\Omega^{(\zeta)}(\alpha,\kvec)=
\frac{3\es}{32\pi^3\sqrt{2 N_c}}}
\nonumber\\
&&\times
\sum_{\lambda'\lambda}\Gamma^{(\zeta)*}_{\lambda' \lambda}(\alpha,\kvec)
\hat{\sigma}^{c\bar{c}N}\Psi^{\gamma(\zeta)}_{\lambda' \lambda}(\alpha,\kvec)
\label{eq:Omega}
\end{eqnarray}
with Eqs.~(\ref{eqn:photon-wL}), (\ref{eqn:photon-wT}) 
for $\Psi^{\gamma(\zeta)}_{\lambda' \lambda}(\alpha,\kvec)$,
and Eq.~(\ref{eqn:A2}) for $\hat{\sigma}^{c\bar{c}N}$.
Eq.~(\ref{eq:ampnew})
provides the basis of our study for the Fermi motion effects:
it is given by the integral of $\Omega^{(\zeta)}(\alpha,\kvec)$,
weighted by the complicated function
$\phi^{*}(\alpha,\kvec)$, whose ($\alpha, \kvec)$-dependence represents the relative motion
between $c$ and $\bar{c}$ due to
the nonperturbative dynamics inside the vector meson.

\subsection{Fermi motion effects in the LLA diffractive amplitude and
comparison with other treatments}

For a nonrelativistic $c\bar{c}$ bound-state,
the WF $\phi^{*}(\alpha,\kvec)$ is sharply peaked at $\alpha = 1/2$ and $\kvec=0$.
Thus, we may conveniently identify the average velocity $v$
of the charm quark in the charmonium rest frame as, e.g.,
\begin{equation}
v^{2} =\frac{3}{2}\left\langle \frac{\kvec^2}{m_{c}^{2}}\right\rangle 
\simeq 12 \left\langle \left(\alpha-\frac{1}{2}\right)^{2}\right\rangle,
\label{eq:v2}
\end{equation}
with
\begin{equation}
\langle \cdots \rangle \equiv 
\frac{\int d\alpha d^2 k_{\perp}\cdots\phi^{*}(\alpha,\kvec)}
{\int d\alpha d^2 k_{\perp}\phi^{*}(\alpha,\kvec)}.
\label{eq:lran}
\end{equation}
In order to see the dependence of ${\rm Im}{\cal M}^{(\zeta)}$
on the velocity $v$,
we may Taylor expand $\Omega^{(\zeta)}(\alpha,\kvec)$ around $\alpha = 1/2$ and $\kvec = 0$
in the integrand of Eq.~(\ref{eq:ampnew}), 
and then perform the integrations over $\alpha$ and $\kvec$
with the weight $\phi^{*}(\alpha,\kvec)$:
the leading term gives the static limit, 
which corresponds to no relative motion of the quarks, i.e.,
the replacement $\phi^{*}(\alpha,\kvec)\rightarrow {\rm const} 
\times\delta(\alpha-1/2)\delta^{(2)}(\kvec)$
in Eq.~(\ref{eq:ampnew}), and the next-to-leading term gives the Fermi motion effects 
of order $v^{2}$. 
Since all realistic nonperturbative models for $\phi^{*}(\alpha,\kvec)$ would give similar values 
for Eq.~(\ref{eq:v2}) as $v^{2} \simeq 0.2 \sim 0.3$, 
the size of the Fermi motion effects in
${\rm Im}{\cal M}^{(\zeta)}$ 
should be essentially determined by
that of the expansion coefficients
in the Taylor expansion of $\Omega^{(\zeta)}(\alpha,\kvec)$.
Namely, the dependence of $\Omega^{(\zeta)}(\alpha,\kvec)$ on $\alpha$ and $\kvec$ 
determines the actual
significance of the Fermi motion corrections.

The ($\alpha, \kvec)$-dependence of Eq.~(\ref{eq:Omega})
from $\Psi^{\gamma(\zeta)}_{\lambda' \lambda}(\alpha,\kvec)$ 
is unambiguously determined to be Eqs.~(\ref{eqn:photon-wL}), (\ref{eqn:photon-wT}) 
in the LLA.
On the other hand, for the other parts $\Gamma^{(\zeta)*}_{\lambda' \lambda}(\alpha,\kvec)$ and $\hat{\sigma}^{c\bar{c}N}$, 
the previous works by RRML and FKS employed
different shapes from our Eqs.~(\ref{eqn:L1})-(\ref{eqn:T2}) and Eq.~(\ref{eqn:A2}).
Because those different shapes could be a major source to influence the Fermi motion corrections,
it is instructive to compare our 
$\Gamma^{(\zeta)*}_{\lambda' \lambda}(\alpha,\kvec)$ and
$\hat{\sigma}^{c\bar{c}N}$ with those of the previous works. 

Now, we discuss about the effective
$c\bar{c} \rightarrow V$ vertex
$\Gamma^{(\zeta)*}_{\lambda' \lambda}(\alpha,\kvec)$.
In the RRML paper \cite{RRML}, their effective vertex reads
\begin{equation}
\Gamma^{(0)*}_{\lambda'\lambda; {\rm RRML}}
=- 2 \delta_{\lambda', -\lambda}
\label{eqn:IID-0}
\end{equation}
for the longitudinal polarization, and 
\begin{eqnarray}
\Gamma_{\lambda'\lambda; {\rm RRML}}^{(\tau)*}
=\left\{ {\begin{array}{lr}
 \displaystyle{
\frac{\sqrt{2}\ m_c \lambda\delta_{\tau,\lambda} }{\alpha(1-\alpha)M_V}} & 
(\lambda'=\lambda)  \\
\\
\displaystyle{\frac{2(\vepvec^{(\tau) *}\cdot\kvec)
(\alpha-\delta_{\tau, \lambda})}{\alpha(1-\alpha)M_V}}
&\;\;\;\; (\lambda'=-\lambda) \\
\end{array}} \right.
\label{eqn:IID-1}
\end{eqnarray}
for the transverse polarization with $\tau = \pm1$.  
To show the relation of the RRML vertex with ours,
we calculate Eq.~(\ref{eqn:GamV}) with the replacement ${\cal R} \rightarrow 1$:
\begin{eqnarray}
\left.\Gamma^{(0)*}_{\lambda'\lambda}\right|_{{\cal R} \rightarrow 1}
&=&- \left(1 + \frac{\kvec^2 + m_{c}^{2}}{\alpha(1-\alpha)M_{V}^{2}}\right) 
\delta_{\lambda', -\lambda},
\label{eqn:IID-new1}\\
\left.\Gamma_{\lambda'\lambda}^{(\pm 1)*}\right|_{{\cal R} \rightarrow 1}
&=&\Gamma_{\lambda'\lambda; {\rm RRML}}^{(\pm 1)*}.
\label{eqn:IID-new2}
\end{eqnarray}
Thus, $\Gamma_{\lambda'\lambda; {\rm RRML}}^{(\zeta)*}$ is given by
$\left. \Gamma_{\lambda'\lambda}^{(\zeta)*}\right|_{{\cal R} \rightarrow 1}$ with imposing the condition,
\begin{equation}
\frac{\kvec^2 + m_{c}^{2}}{\alpha}+
\frac{\kvec^2 + m_{c}^{2}}{1-\alpha}
= M_{V}^{2},
\label{eq:onsh}
\end{equation}
which means that the corresponding $c\bar{c} \rightarrow V$ vertex is on the energy shell:
we note that the Lepage-Brodsky spinors $u_{\lambda}(k)$, $v_{\lambda}(k)$
are the functions of only the components $(k^{+}, \kvec)$ of the momentum $k$ \cite{LB},
so that the vertex defined as Eq.~(\ref{eqn:GamV})
obeys the conservation for the ``three-momentum'' $(k^{+}, \kvec)$ only.
Namely, the sum of the three-momenta of quark and antiquark
equals the three-momentum of the vector meson, $(q^{+}+\Delta^{+}, \qvec+\Delvec)= (q^{+}, \0vec)$,
but the conservation of the ``minus'' component of the momentum, expressed as Eq.~(\ref{eq:onsh}), needs not be satisfied.
(See also Ref.~\cite{RC}. In fact, the situation is similar to that for the spinor matrix elements
of the photon WF (\ref{eq:Psig}), that represent the $\gamma \rightarrow c\bar{c}$ vertex.)
Comparing Eqs.~(\ref{eqn:IID-0}), (\ref{eqn:IID-1}) 
with Eqs.~(\ref{eqn:IID-new1}), (\ref{eqn:IID-new2}),
we see that the Fermi motion effects of order $v^{2}$ would be modified if 
the condition (\ref{eq:onsh}) were imposed by hand.

Also, comparing our vertex~(\ref{eqn:L1})-(\ref{eqn:T2})
with Eqs.~(\ref{eqn:IID-new1}), (\ref{eqn:IID-new2}),
we find that the replacement ${\cal R} \rightarrow 1$
would modify the Fermi motion effects of the order $v^{2}$. 
Apparently, the spin structure of Eqs.~(\ref{eqn:IID-new1}), (\ref{eqn:IID-new2})
would be contaminated by that for the $D$-wave $c \bar{c}$ states,
so that these vertices cannot be combined with nonrelativistic models 
for the WF $\phi^{*}(\alpha, \kvec)$. 

On the other hand, FKS also considered their effective $c\bar{c} \rightarrow V$ vertex
on the energy shell with Eq.~(\ref{eq:onsh}),
but tried to include the spin structure for a pure $^3 S_1$ $c \bar{c}$-state \cite{FKS}.
They employed the projector, defined 
in the helicity representation as (see Eq.~(\ref{eqn:GamV}))
\begin{equation}
{\cal R}_{\rm FKS} = \delta_{\zeta, 0}\delta_{\lambda',-\lambda}
+ \left(\delta_{\zeta, 1}+ \delta_{\zeta, -1}\right)\delta_{\lambda',\lambda},
\label{eq:RFKS}
\end{equation}
so that the FKS vertex reads
\begin{eqnarray}
\Gamma_{\lambda'\lambda; {\rm FKS}}^{(0)*}&= &\Gamma_{\lambda'\lambda; {\rm RRML}}^{(0)*},
\label{eq:FKSL}
\\
\Gamma_{\lambda'\lambda; {\rm FKS}}^{(\pm1)*}&=& 
\delta_{\lambda',\lambda}\Gamma_{\lambda'\lambda; {\rm RRML}}^{(\pm 1)*}.
\label{eq:FKST}
\end{eqnarray}
The FKS projector (\ref{eq:RFKS}) would give the exact spin structure for the triplet states,
if it acted on the helicities of the two-component Pauli spinors.
However, when it acts on the helicities of the {\it Dirac} spinors as in the treatment by FKS,
it is not the correct projector to ensure the spin-triplet states;
the corresponding error 
modifies the Fermi motion effects of order $v^{2}$ 
(compare Eqs.~(\ref{eq:FKSL}), (\ref{eq:FKST}) 
with Eqs.~(\ref{eqn:L1})-(\ref{eqn:T2})).

Now, we recognize that our vertex (\ref{eqn:GamV}) is different
from that of RRML as well as that of FKS in two points: 
the correct spin structure with the projector ${\cal R}$ of Eq.~(\ref{eqn:projection})
and the off-shellness without the condition (\ref{eq:onsh}).
Both of these points modify ${\rm Im}{\cal M}^{(\zeta)}$ 
of Eq.~(\ref{eq:ampnew}) at the order $v^{2}$
and lead to the additional Fermi motion corrections compared with those of
RRML or FKS.

There is, still, another ``new'' Fermi motion correction from
$\hat{\sigma}^{c\bar{c}N}$ of Eq.~(\ref{eqn:A2}):
in almost all previous works including RRML and FKS, 
$\tilde{x}$ appearing in $\hat{\sigma}^{c\bar{c}N}$ is replaced as
\begin{equation}
\tilde{x}\rightarrow  x=\frac{Q^{2} +M_{V}^{2}}{\es} .
\label{eq:x}
\end{equation}
Because this $x$ corresponds to the static limit of the on-energy-shell value of $\tilde{x}$
(\ref{eq:tilx}), the use of $\tilde{x}$ 
with the explicit $\kvec$- and $\alpha$-dependence
generates the $O(v^2)$ modification of ${\rm Im}{\cal M}^{(\zeta)}$.

Before ending this subsection,
we mention the connection of our projector (\ref{eqn:projection})
with the operator used in Ref.~\cite{IIvanov} to ensure the $S$-wave 
$c\bar{c}$ states.
When we impose the on-energy-shell condition (\ref{eq:onsh}) by hand,
Eq.~(\ref{eqn:GamV}) reduces to
\begin{equation}
\frac{M_{V}+2m_{c}}{2M_V^{2} \sqrt{\alpha(1-\alpha)}}
\bar{v}_{\lambda}(q-k)\gamma_{\mu}
{e_{\nu}^{(\zeta)}}^{*} {\cal S}^{\mu \nu}
u_{\lambda'}(k),
\label{eqn:Iva}
\end{equation}
with
${\cal S}^{\mu \nu} = g^{\mu \nu} - 2 k^{\mu}k^{\nu}/[m_{c}(M_{V} + 2m_{c})]$,
using the Dirac equation for the spinors.
This operator ${\cal S}^{\mu \nu}$ was used in Ref.~\cite{IIvanov}. 
This result shows that ${\cal S}^{\mu \nu}$
leads to the $O(v^{2})$ error, and it is not a projection operator,
i.e., ${\cal S}^{\mu \nu}{\cal S}_{\nu \rho}\neq
{\cal S}^{\mu}_{\rho}$.

\subsection{Production cross section with Fermi motion effects}

We are now in a position to present our final formula 
for the forward differential cross section of the 
diffractive heavy vector-meson production including all the Fermi motion effects discussed above,
and also for the corresponding total production cross section.
Substituting Eqs.~(\ref{eqn:photon-wL}), (\ref{eqn:photon-wT}), (\ref{eqn:A2}),
(\ref{eqn:L1})-(\ref{eqn:T2}) into Eq.~(\ref{eq:Omega}), we get
\begin{widetext}
\begin{eqnarray}
\Omega^{(0)}(\alpha, \kvec) &=&
\frac{ee_c \nutilde Q}{4\pi \sqrt{2N_c}}
\alpha(1-\alpha)\alpha_s(Q_{\rm eff}^2)\tilde{x}G(\tilde{x},Q_{\rm eff}^2)
\frac{\Qbar^2 -\kvec^2}{\left(\kvec^2+\Qbar^2\right)^3}
\left[1+\frac{\kvec^2+m_c(M_V +m_c)}
{\alpha(1-\alpha)M_V^2}\right],
\label{eqn:7-1}\\
\Omega^{(\pm 1)}(\alpha, \kvec)
&=&
\frac{ee_c\nutilde}{8\pi\sqrt{2N_c}M_V^2}
\frac{\alpha_s(Q_{\rm eff}^2)\tilde{x}G(\tilde{x},Q_{\rm eff}^2)}
{\alpha(1-\alpha)\left(\kvec^2+\Qbar^2\right)^3}        
\left[ m_c \left\{ (M_V +m_c)m_c + \alpha(1-\alpha)M_V^2 \right\}
\left( \kvec^2 -\Qbar^{2}   \right)\right.
\nonumber\\
&-&\left.  2 \left\{ \left[ \alpha^2+(1-\alpha)^2 \right] M_V +m_c \right\}
\kvec^2 \Qbar^2 \right],
\label{eqn:7-2}
\end{eqnarray}
\end{widetext}
for the longitudinal and transverse polarizations, respectively. 
Here we have omitted the contributions that vanish when integrating over 
the angle of $\kvec$ in Eq.~(\ref{eq:ampnew}).
The forward differential cross section is given as
\begin{eqnarray}
\left.\frac{d\sigma(\gamma^{(*)} N \rightarrow V N)}{dt}\right|_{t=0}
=\frac{1}{16\pi W^4}\,\sum_{\zeta}|{\cal M}^{(\zeta)}|^2 .
\label{eqn:sigma0}
\end{eqnarray}
Following Refs.~\cite{BFGMS,FKS,RRML}, 
we can calculate ${\rm Re}{\cal M}^{(\zeta)}$ perturbatively
from ${\rm Im}{\cal M}^{(\zeta)}$,
using ${\rm Re}{\cal M}^{(\zeta)}/{\rm Im}{\cal M}^{(\zeta)}=\rho$ with 
\begin{eqnarray}
\rho \approx \frac{\pi}{2}
\frac{\partial \ln \left(xG(x,Q_{\rm eff}^2)\right)}{\partial\ \ln x}.
\label{eqn:1-2} 
\end{eqnarray}
The forward differential cross section (\ref{eqn:sigma0})
is now given as
\begin{equation}
\left.\frac{d\sigma(\gamma^{(*)} N \rightarrow V N)}{dt}\right|_{t=0}
\!\!\!\!=\frac{1}{16\pi W^4}\sum_{\zeta} (\mbox{Im}{\cal M}^{(\zeta)})^2
\left(1+\rho^2\right).
\label{eqn:sigma}
\end{equation}
(The actual contribution from the real part is 
found to be less than 10\% at HERA energies.)
Combining the result (\ref{eqn:sigma}) with the $t$-dependence of the 
differential cross section observed in experiment, $d\sigma/dt \propto \exp(B_{V}t)$,
with a constant diffractive slope $B_{V}$ \cite{FKS},   
we can calculate the total cross section for the diffractive vector-meson production,
\begin{equation}
\sigma (\gamma^{(*)} N \rightarrow V N) = \frac{1}{B_{V}}
\left. \frac{d\sigma(\gamma^{(*)} N \rightarrow V N)}{dt} \right|_{t=0} .
\label{eqn:sigmatot}
\end{equation}

To obtain ${\rm Im}{\cal M}^{(\zeta)}$ of Eq.~(\ref{eqn:sigma}),
we use Eqs.~(\ref{eqn:7-1}), (\ref{eqn:7-2}) in Eq.~(\ref{eq:ampnew}).
In Sec.~II C, we explained the Fermi motion effects by Taylor expanding 
$\Omega^{(\zeta)}(\alpha, \kvec)$ in the integrand.
However, in the numerical calculation presented below, it is more convenient
to evaluate directly the convolution integrals over $\alpha$ and $\kvec$ 
without recourse to the Taylor expansion.
The nonperturbative part of the light-cone WF, $\phi^{*}(\alpha, \kvec)$,
is constructed from 
a $S$-wave solution of the Schr\"{o}dinger equation with a realistic potential
between $c$ and $\bar{c}$.
We use 
the Cornell potential consisted of ``Coulomb plus linear potential'', 
$V(r) = - \kappa/r + r/a^{2}$, with  
the parameters $\kappa$ and $a$ chosen to reproduce the masses and the decay properties of 
the charmonia \cite{QR}.
(The use of the other QCD-inspired potentials 
modifies the production cross section by at most 10$\%$.)
We follow the procedure of FKS \cite{FKS},
in order to derive $\phi^{*}(\alpha, \kvec)$ 
with the light-cone variables $\alpha$, $\kvec$ 
from the solution of the conventional 
``equal-time'' Schr\"{o}dinger equation, $\phi_{NR}(|\vec{k}|)$,
where $\vec{k}$ denotes the (three-dimensional) variable 
for the space components of the momentum $k$:
we employ the kinematical 
identification of the Sudakov variable $\alpha$,
which denotes the fraction of the ``plus'' component of the meson's momentum
carried by the $c$-quark, with
\begin{eqnarray}
\alpha = \frac{1}{2} \left( 1+ \frac{k_z}{\sqrt{\vec k^2 + m_c^2}} \right) .
\label{eqn:3momentum}
\end{eqnarray}
This relation, together with the conservation of the overall normalization of the WF,
$\int d\alpha d^{2}k_{\perp}|\phi (\alpha, \kvec)|^{2}/2 = \int d^{3}k |\phi_{NR}(|\vec{k}|)|^2$,
allows us to express the light-cone WF in terms of the nonrelativistic WF as
\begin{eqnarray} 
\phi (\alpha, \kvec) &=&  
\left({\frac{\kvec^2 + m_c^2 }{4 [\alpha (1-\alpha )]^3}}\right)^{1/4} 
\nonumber\\
\times&&\!\!\!\!\!\!\!\!\!\!
\phi_{NR} \left( |\vec k| = \sqrt{\frac{\kvec ^2+(2\alpha - 1)^2 m_c^2}{4 \alpha (1-\alpha)}}
\right) .
\label{eqn:scalar-wf}
\end{eqnarray}

The obtained WF $\phi(\alpha,\kvec)$ is in fact peaked at $\alpha=1/2$
and $\kvec = 0$. When we take the static limit
$\phi(\alpha,\kvec)\rightarrow {\rm const} \times\delta(\alpha-1/2)\delta^{(2)}(\kvec)$ and 
impose the condition (\ref{eq:onsh}), 
it is easy to see that the differential cross section (\ref{eqn:sigma}) coincides
with the old result derived by Ryskin \cite{Ryskin},
\begin{equation}
\left.\frac{d\sigma}{dt}\right|_{t=0}\!\!\!\!
\rightarrow \frac{\Gamma_{ee}M_{V}^3 \pi^3}{48 \alpha_{\rm em} Q_{\rm eff}^{8}}
\left[\alpha_s(Q_{\rm eff}^2)xG(x,Q_{\rm eff}^2)\right]^{2}
\left(1 + \frac{Q^{2}}{M_{V}^{2}}\right),
\label{eq:rys}
\end{equation}
up to the small contribution due to $\rho$.
Here, $\Gamma_{ee}$ stands for the decay width of the vector meson 
into an $e^{+} e^{-}$ pair. In the last parenthesis, the first and the second terms
correspond to the production with transversely and longitudinally polarized photons, 
$\sigma_{T}$ and $\sigma_{L}$, respectively. 
As we will demonstrate in the next section,
the dominant effects due to the Fermi motion corrections with the full WF (\ref{eqn:scalar-wf})
modify these two contributions by the different factors 
that could in principle depend on $Q^{2}$ as well as $W$.

On the other hand, 
if we formally suppose $Q^{2} \gg m_{c}^{2}$ in Eq.~(\ref{eqn:sigma})
and impose the condition (\ref{eq:onsh}), the result reproduces
the LLA formula of Brodsky {\it et al.} \cite{BFGMS} 
for the hard diffractive production.  
In this case, $\Qbar^2 = m_{c}^{2} + \alpha(1-\alpha)Q^{2}$ 
becomes much larger than $\kvec^2$ in Eqs.~(\ref{eqn:7-1}), (\ref{eqn:7-2}), except the end-point region of the $\alpha$-integral,
so that the $\alpha$-dependence of the corresponding integrand plays a dominant role
in the cross section compared with the $\kvec$-dependence; 
this situation corresponds to the light-cone dominance
and the $\kvec$-dependence is only responsible 
for the small higher-twist ($\sim 1/Q^{2}$) corrections.
However, in the region $Q^{2} \lesssim m_{c}^{2}$ discussed below,
the $\kvec$-dependence is as important as 
the $\alpha$-dependence in Eqs.~(\ref{eqn:7-1}), (\ref{eqn:7-2}).
In particular, for the photoproduction $Q^2=0$, 
the transverse quark motion represented by the $\kvec$-dependence could play
a more dominant role in the Fermi motion effects.  
To show the quantitative role of the transverse quark motion, 
we will present the results of the cross sections 
in the ``static limit for the transverse motion''
with the replacement in Eq.~(\ref{eq:ampnew}),
\begin{equation}
\phi(\alpha, \kvec) \rightarrow \varphi(\alpha)\delta^{(2)}(\kvec),
\label{eq:static}
\end{equation}
where $\varphi(\alpha) = \int d^{2}k_{\perp} \phi(\alpha, \kvec)$,
and compare with the results using the full WF (\ref{eqn:scalar-wf}).
We will demonstrate that the significant effect from the transverse quark motion
is a novel feature in the production of the heavy mesons.

\section{The $J/\psi$ and $\psi'$ production cross section and comparison with data}

In this section, we present the numerical results, using our formulae
(\ref{eqn:sigma}) and (\ref{eqn:sigmatot})
with Eqs.~(\ref{eq:ampnew}), (\ref{eqn:7-1}), (\ref{eqn:7-2}), and (\ref{eqn:scalar-wf}).
We show the cross section for the diffractive photo- and electroproductions
of the heavy vector-mesons $V= J/\psi, \psi'$, 
and discuss the roles of the Fermi motion effects
in detail.
We use $m_c = 1.5$ GeV for the charm quark mass, and 
$M_{J/\psi}= 3.10$ GeV and $M_{\psi'} = 3.69$ GeV
for the mass of the vector mesons.
For the slope parameter $B_{V}$ of Eq.~(\ref{eqn:sigmatot}),
we adopt $B_{J/\psi}=4.44$ GeV$^{-2}$ for $J/\psi$ 
and $B_{\psi'}=4.31$ GeV$^{-2}$ for $\psi '$,
that are extracted from the experiments \cite{H100,H102,ZEUS02}.

For the gluon distribution function $G(\tilde{x},Q^2_{\rm eff})$ 
of Eqs.~(\ref{eqn:7-1}), (\ref{eqn:7-2}),
we employ Gl\"{u}ck-Reya-Vogt (GRV) parametrization \cite{GRV95}.
Besides uncertainties in the knowledge of $G(\tilde{x},Q^2_{\rm eff})$,
we cannot determine the scale $Q^2_{\rm eff}$,
which enters into $G(\tilde{x},Q^2_{\rm eff})$ 
as well as the strong coupling constant $\alpha_s(Q^2_{\rm eff})$,
unambiguously within the accuracy of the LLA.
We also note that, within the LLA, the use of the NLO fits to $G(\tilde{x},Q^2_{\rm eff})$
is not fully consistent.
In fact, in the previous works, the various modifications of $Q^2_{\rm eff}$
were employed to try to simulate the important effects 
beyond the LLA \cite{FKS,RRML,IIvanov,NNPZZ}.
In the present work, we do not pursue such ``schemes'' to fix the scale $Q^2_{\rm eff}$.
In addition to the results with the ``standard'' scale $Q^2_{\rm eff}=(Q^2 + M_V^2)/4$
in Eqs.~(\ref{eqn:7-1}) and (\ref{eqn:7-2}),
we will present the results with the replacement
\begin{equation}
Q^2_{\rm eff} \rightarrow 2 Q^2_{\rm eff},
\label{eq:rescale}
\end{equation}
and 
thereby we demonstrate how much our results would
be modified by the corresponding ambiguity.
We note that actually the replacement (\ref{eq:rescale}) 
gives very similar effects to those of the ``$Q^2$ rescaling''
employed by FKS, as one can read off from Fig.9 in the first paper of Ref.~\cite{FKS}.

\subsection{Fermi motion effects and the $\sigma_{L}/\sigma_{T}$ ratio}
\vspace*{0.3cm}
\begin{figure*}[htb]
\setlength{\unitlength}{1cm}
\begin{minipage}[t]{7.7cm}
\begin{picture}(6.5,6.5)
\hspace*{-0.5cm}
\psfig{file=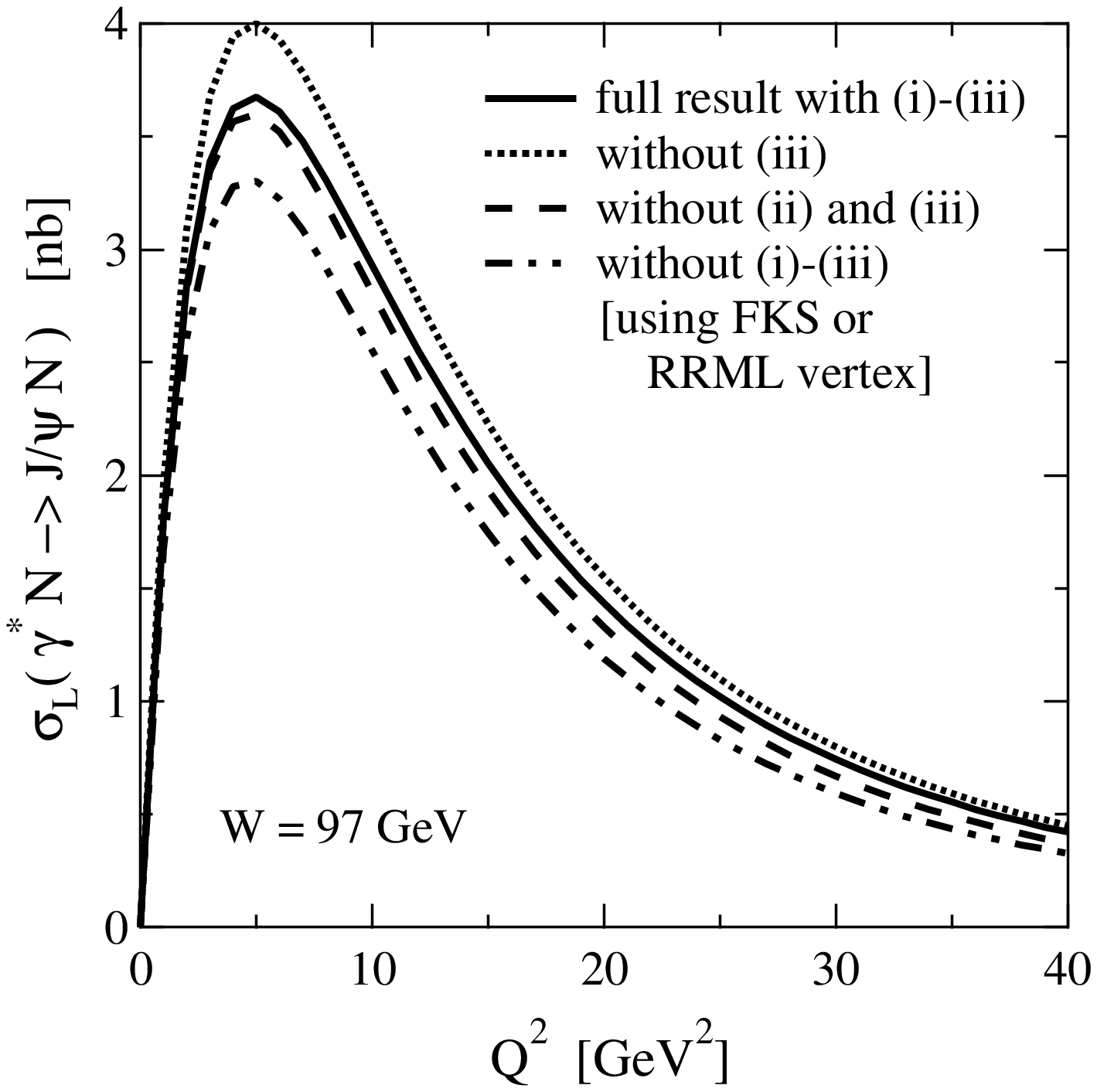,width=7.5cm,height=7.5cm}
\end{picture}\par
\caption{The cross section for the diffractive $J/\psi$ production
with longitudinally polarized photon, as a function of $Q^{2}$ at $W=97$ GeV.
The solid line shows our full result (\ref{eqn:sigmaL}),
including the new Fermi motion effects due to (i)-(iii).
The dotted line is the result excluding (iii),
and the dashed line is the one excluding (ii) and (iii).
The two-dot-dashed line shows the result without (i)-(iii),
using the RRML vertex (\ref{eqn:IID-0}) or the FKS vertex (\ref{eq:FKSL}).
}
\label{fig:L1s_Qdep_model}
\end{minipage}
\hspace*{0.7cm}
\begin{minipage}[t]{7.7cm}
\begin{picture}(6.5,6.5)
\hspace*{-0.5cm}
\psfig{file=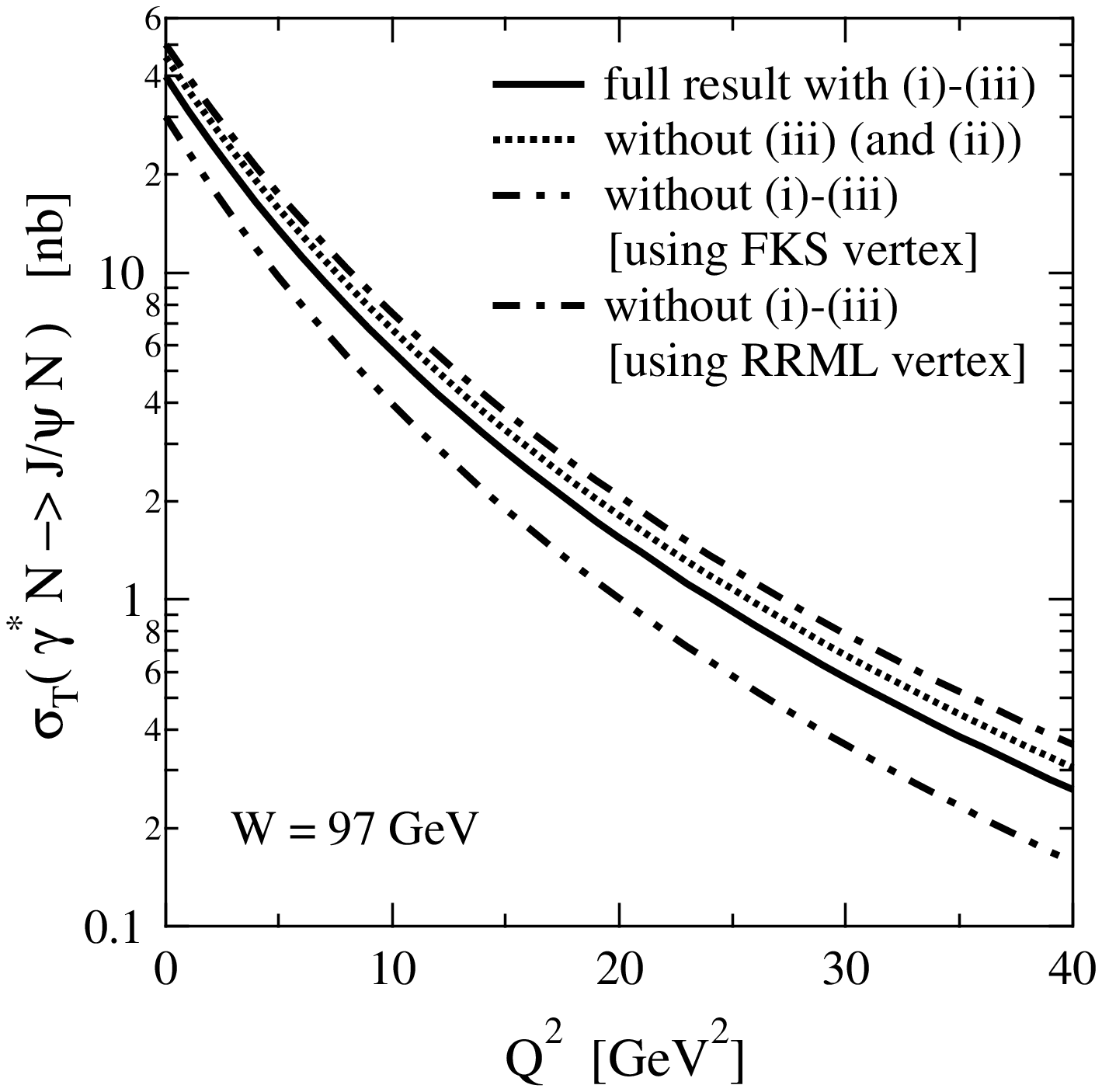,width=7.5cm,height=7.5cm}
\end{picture}\par
\caption{The cross section for the diffractive $J/\psi$ production
with transversely polarized photon, as a function of $Q^{2}$ at $W=97$ GeV.
The solid line shows our full result (\ref{eqn:sigmaT}),
including the new Fermi motion effects due to (i)-(iii).
The dotted line shows the result excluding (iii), as well as excluding (ii) and (iii).
The dot-dashed and the two-dot-dashed lines show the result without (i)-(iii),
using the RRML vertex (\ref{eqn:IID-1}) and the FKS vertex (\ref{eq:FKST}), respectively.
}
\label{fig:T1s_Qdep_model}
\end{minipage}
\end{figure*}
\begin{figure}[h]
\vspace*{0cm}
\begin{center}
\hspace*{0cm}
\psfig{file=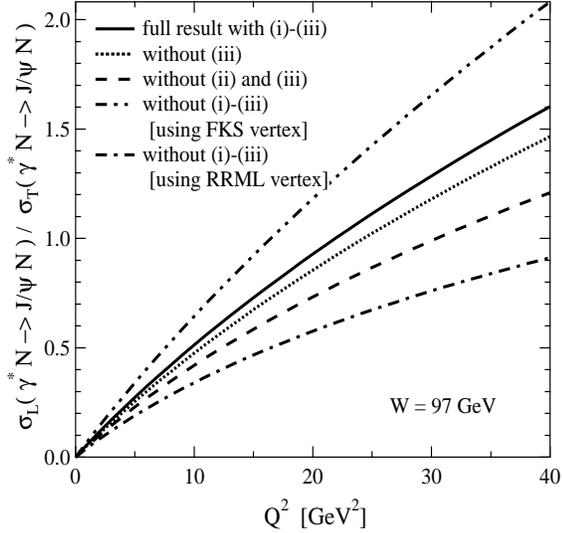,width=8cm,height=7.5cm}
\end{center}
\vspace{0cm}
\caption{The ratio of the cross sections for the diffractive $J/\psi$ production
with longitudinally and transversely polarized photons, 
as a function of $Q^{2}$ at $W=97$ GeV.
Each curve is obtained by taking the ratio of the corresponding results
of Figs.~\ref{fig:L1s_Qdep_model} and \ref{fig:T1s_Qdep_model},
and the lines have the same meaning as in those figures.
}
\label{fig:LT1s_Qdep_model}
\end{figure}
%
As pointed out in Sec. II,
the new Fermi motion effects in the LLA, compared with the previous works,
originate from three points:
\begin{enumerate}
\renewcommand{\labelenumi}{(\roman{enumi})}
\item 
The spin WF for the $^{3}S_{1}$ $c\bar{c}$ state 
with the projector ${\cal R}$ (\ref{eqn:projection}).
\item 
The off-shell $c\bar{c} \rightarrow V$ vertex without using 
the condition (\ref{eq:onsh}).
\item 
The modification of the gluon's longitudinal momentum fraction probed by the process,
as Eq.~(\ref{eq:tilx}), due to the coupling 
with the $c\bar{c}$ pair in internal relative motion.
\end{enumerate}

In order to understand the roles of these new effects,
it is useful to compare our vertex involving (i) and (ii)
with those of RRML and FKS:
our vertex, Eqs.~(\ref{eqn:L1})-(\ref{eqn:T2}), has quite different
$\alpha$ and $\kvec$ behavior 
from the RRML vertex (\ref{eqn:IID-0}), (\ref{eqn:IID-1})
as well as the FKS vertex (\ref{eq:FKSL}), (\ref{eq:FKST}),
for each $(\lambda, \lambda')$ and for both longitudinally 
and transversely polarized vector-mesons.
Namely, (i) combined with (ii) is expected to produce
strongly helicity-dependent Fermi motion effects.
We further note that (ii) itself gives the helicity-dependent effects:
by imposing the condition (\ref{eq:onsh}) on our vertex (\ref{eqn:GamV}),
we get Eq.~(\ref{eqn:Iva}). Apparently, this result corresponds
to our vertex (\ref{eqn:L1})-(\ref{eqn:T2}) with the term $\kvec^2$ eliminated by using
Eq.~(\ref{eq:onsh}), and this affects $\Gamma_{-\lambda\lambda}^{(0)*}$ of Eq.~(\ref{eqn:L2})
only.
On the other hand, (iii) is not expected to produce
strong helicity-dependence: the effect of (iii) acts   
similarly on both longitudinal and transverse polarizations
through the gluon distribution $\tilde{x}G(\tilde{x},Q^2_{\rm eff})$
in Eqs.~(\ref{eqn:7-1}) and (\ref{eqn:7-2}).

To investigate the helicity-dependence of our new Fermi motion effects in detail,
we compute the $J/\psi$ production cross section with longitudinally and transversely polarized 
photons (see Eqs.~(\ref{eqn:sigma}), (\ref{eqn:sigmatot})),
\begin{eqnarray}
\lefteqn{\sigma_{L}(\gamma^{(*)} N \rightarrow J/\psi\ N)}
\nonumber\\
&&=\frac{1}{16\pi W^4 B_{J/\psi}}(\mbox{Im}{\cal M}^{(0)})^2
\left(1+\rho^2\right),
\label{eqn:sigmaL}\\
\lefteqn{\sigma_{T}(\gamma^{(*)} N \rightarrow J/\psi\ N)}
\nonumber\\
&&= \frac{1}{16\pi W^4 B_{J/\psi}}\sum_{\zeta=\pm 1} (\mbox{Im}{\cal M}^{(\zeta)})^2
\left(1+\rho^2\right),
\label{eqn:sigmaT}
\end{eqnarray}
with Eqs.~(\ref{eq:ampnew}), (\ref{eqn:7-1}), (\ref{eqn:7-2}), and (\ref{eqn:scalar-wf}).
%
\vspace*{0.3cm}
\begin{figure*}[htb]
\setlength{\unitlength}{1cm}
\begin{minipage}[t]{7.7cm}
\begin{picture}(6.5,6.5)
\hspace*{-0.5cm}
\psfig{file=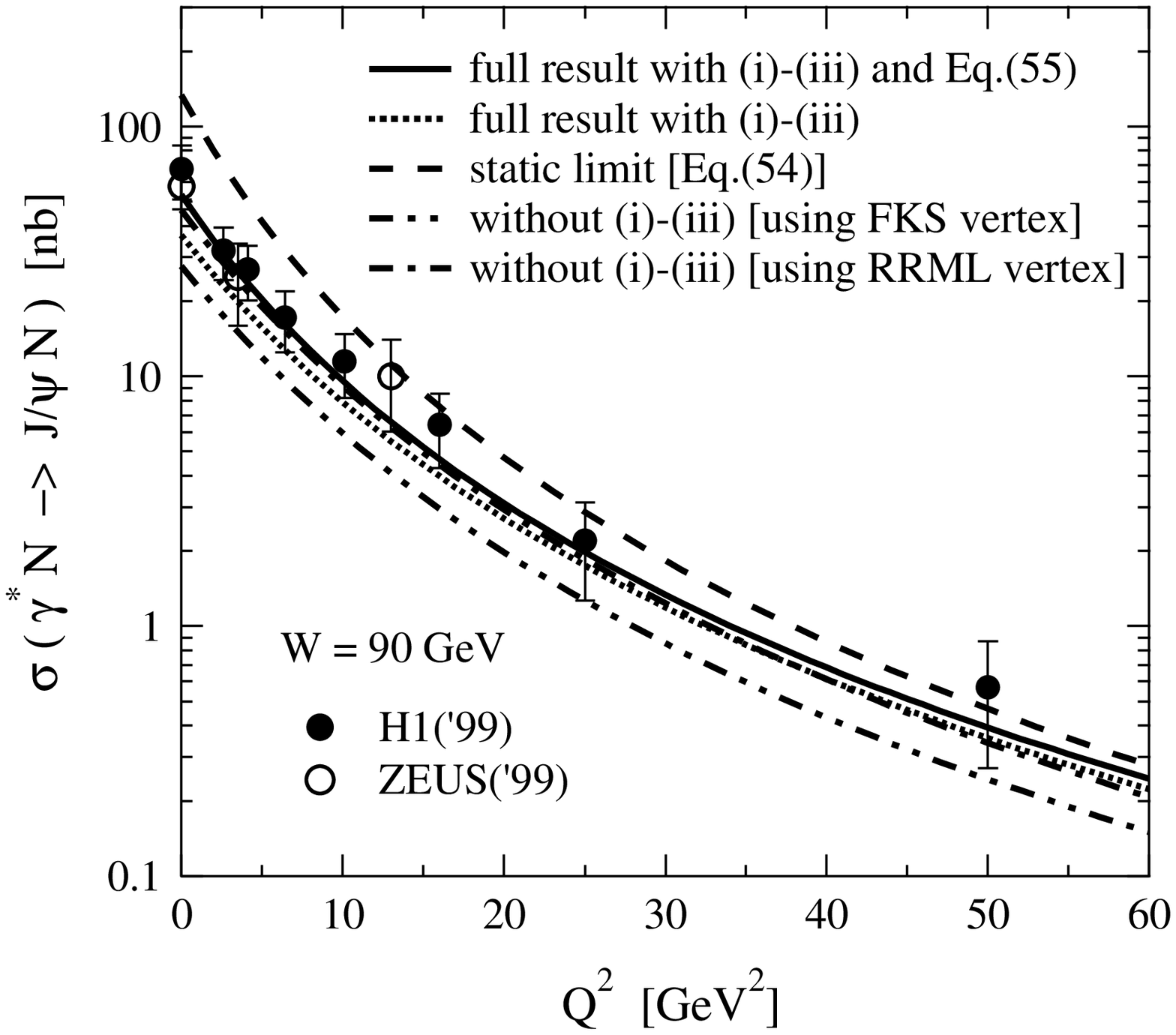,width=8cm,height=7.5cm}
\end{picture}\par
\caption{The total cross section for the electroproduction,
$\gamma^{*}N\rightarrow J/\psi\ N$, as a function of $Q^2$
at $W=90$ GeV, in comparison with the HERA data \cite{H199,ZEUS99}. 
The solid and the dotted lines show our full result (\ref{eqn:sigmatot})
with and without the replacement (\ref{eq:rescale}), respectively.
The dot-dashed and the two-dot-dashed lines show the results 
without the effects due to (i)-(iii), using
the RRML vertex (\ref{eqn:IID-0}), (\ref{eqn:IID-1}) 
and the FKS vertex (\ref{eq:FKSL}), (\ref{eq:FKST}), respectively.
The dashed curve is obtained
from the dotted curve with the replacement (\ref{eq:static}).
}
\label{fig:1s_Qdep}
\end{minipage}
\hspace*{0.7cm}
\begin{minipage}[t]{7.7cm}
\begin{picture}(6.5,6.5)
\hspace*{-0.5cm}
\psfig{file=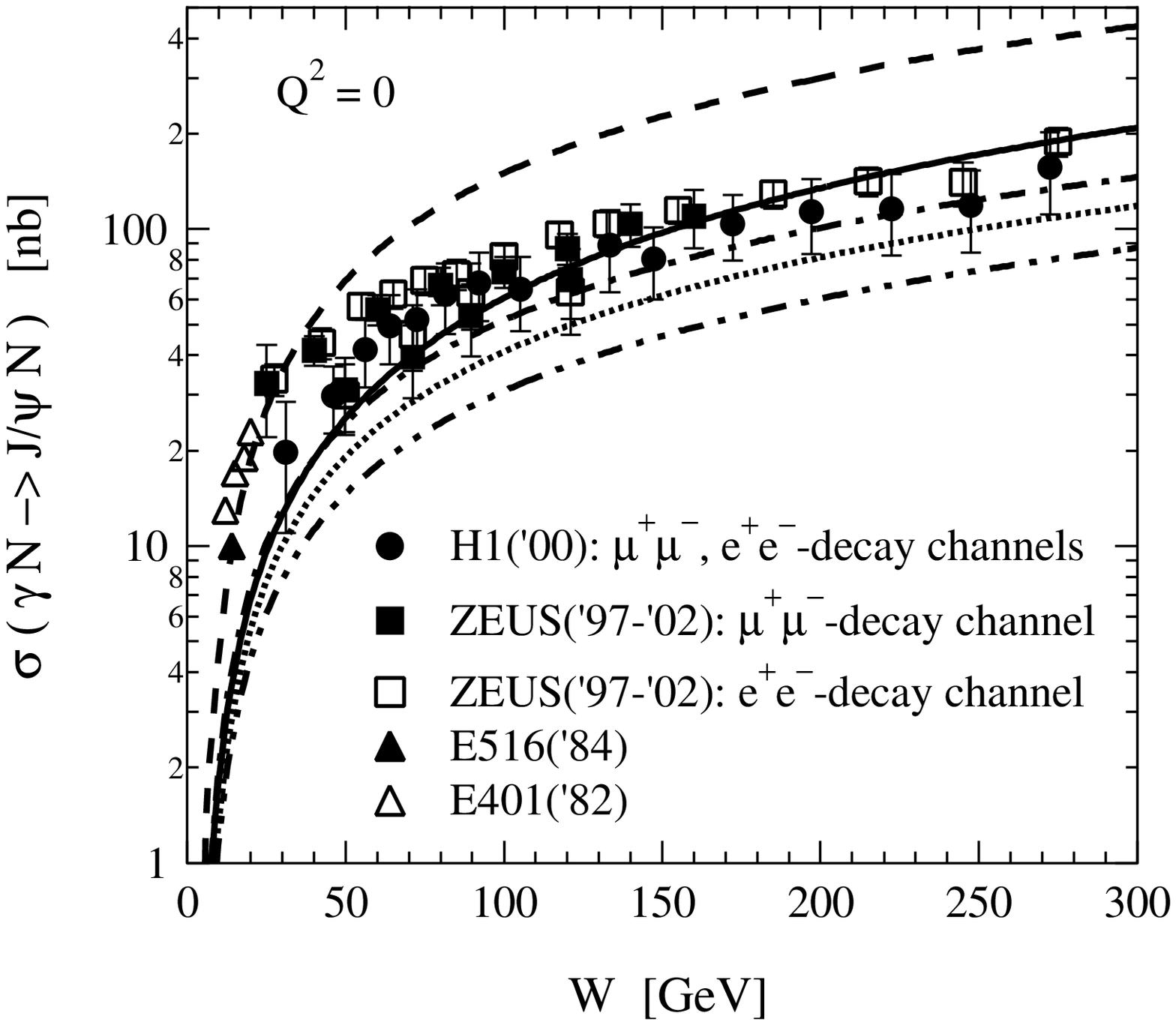,width=8cm,height=7.5cm}
\end{picture}\par
\caption{The total cross section for the photoproduction,
$\gamma N\rightarrow J/\psi\ N$, as a function of $W$.  
The lines have the same meaning as in Fig.~\ref{fig:1s_Qdep}.
Experimental data points from
the H1\cite{H100}, ZEUS\cite{ZEUS97,ZEUS02}, 
E401\cite{E401a}, and E516\cite{E516} experiments.
The HERA data were measured in two leptonic decay channels, 
$J/\psi\rightarrow \mu^+\mu^-$, and $J/\psi\rightarrow e^+e^-$. 
}
\label{fig:1s_Wdep}
\end{minipage}
\end{figure*}
%
%

%
In Fig.~\ref{fig:L1s_Qdep_model},
we show $\sigma_{L}(\gamma^{*} N \rightarrow J/\psi\ N)$ 
as a function of $Q^{2}$
with the $\gamma^{*}$-$N$ center-of-mass energy fixed to be $W=97$ GeV.
The full result (\ref{eqn:sigmaL}) 
is shown by the solid curve, which includes 
all the above-mentioned effects (i), (ii), and (iii).
The dotted curve shows the result without (iii), 
i.e., the result of Eq.~(\ref{eqn:sigmaL}) using Eq.~(\ref{eqn:7-1})
with the replacement (\ref{eq:x}).
The dashed curve shows the result without (ii) and (iii).
This is obtained from the result of the dotted curve 
by replacing the $c\bar{c} \rightarrow V$ vertex by Eq.~(\ref{eqn:Iva}).
By the two-dot-dashed curve, we also show the result 
without (i)-(iii): this is obtained from the result of the dashed curve
by replacing the corresponding $c\bar{c} \rightarrow V$ vertex (\ref{eqn:Iva})
by the RRML vertex (\ref{eqn:IID-0}), 
which coincides with the FKS vertex for the longitudinal polarization 
(see Eq.~(\ref{eq:FKSL})).
The dotted curve is largely enhanced due to the effects of (i) and (ii),
in comparison with the two-dot-dashed curve.
Comparing the dotted and the solid curves, we see that (iii) gives the suppression.
We note that the enhancement due to (ii) is almost canceled
by (iii) in this channel (compare the solid and the dashed curves).
Still, our full result shown by the solid curve is larger than the two-dot-dashed curve, 
demonstrating an important role of the Fermi motion effect by (i).

The results for $\sigma_{T}(\gamma^{*} N \rightarrow J/\psi\ N)$ are  
shown similarly in Fig.~\ref{fig:T1s_Qdep_model},
as a function of $Q^{2}$ at $W=97$ GeV.
The full result (\ref{eqn:sigmaT}) involving (i)-(iii)
is shown by the solid curve.
The dotted curve shows the result (\ref{eqn:sigmaT}) with the replacement (\ref{eq:x}),
corresponding to the case without (iii). 
We note that the dotted curve also corresponds to the case without (ii) and (iii),
because our vertex (\ref{eqn:T1}), (\ref{eqn:T2})
coincide with the vertex (\ref{eqn:Iva}) for the transverse polarizations.
By the dot-dashed and the two-dot-dashed curves, we show the results 
without (i)-(iii): the dot-dashed (two-dot-dashed) curve 
is obtained from the result of the dotted curve
by replacing the corresponding $c\bar{c} \rightarrow V$ vertex 
by the RRML vertex (\ref{eqn:IID-1}) (FKS vertex (\ref{eq:FKST})).
Comparing the solid curve with the dotted curve, we see that 
(iii) gives the suppression, 
similarly to the case of $\sigma_L$ of Fig.~\ref{fig:L1s_Qdep_model}.
From the comparison of the dotted curve with the dot-dashed curve,
we may say that the effect due to (i) gives rise to the
suppression, contrary to the case of $\sigma_L$; recall that 
the RRML vertex does not take into account
the projector ${\cal R}$ for the $S$-wave $c\bar{c}$ state at all,
i.e., ${\cal R} \rightarrow 1$.
We note that the two-dot-dashed curve using the FKS vertex is much more suppressed than
the dotted curve. This indicates that the FKS projector (\ref{eq:RFKS}),
which has an oversimplified structure 
omitting the $\lambda'=-\lambda$ component of the vertex as in Eq.~(\ref{eq:FKST}),
overestimates the effect of (i) due to the spin WF.

To summarize the helicity-dependence of the Fermi motion effects due to (i)-(iii):
the effect due to (i) enhances $\sigma_{L}$, while it suppresses $\sigma_{T}$;
the effect due to (ii) enhances $\sigma_{L}$, and it does not change $\sigma_{T}$; 
the effect due to (iii) suppresses both $\sigma_{L}$ and $\sigma_{T}$.
We also note 
that the ``oversimplified'' FKS projector (\ref{eq:RFKS}) significantly
overestimates the suppression of $\sigma_{T}$ due to (i).

Those strong helicity-dependence from (i), (ii)
gives rise to notable behavior of the ratio 
$\sigma_L (\gamma^{*} N \rightarrow J/\psi\ N)/\sigma_T (\gamma^{*} N 
\rightarrow J/\psi\ N)$,
which is shown in Fig.~\ref{fig:LT1s_Qdep_model}.
Here each curve is obtained by taking the ratio of the results expressed 
by the corresponding curves in
Figs.~\ref{fig:L1s_Qdep_model} and \ref{fig:T1s_Qdep_model}.
Note that, if we plotted the dashed curve in Fig.~\ref{fig:T1s_Qdep_model} 
(the dot-dashed curve in Fig.~\ref{fig:L1s_Qdep_model}),
it would completely coincide with the dotted (two-dot-dashed) curve. 
The dashed curve is enhanced in comparison with the dot-dashed curve 
due to the ``combined'' effect of (i)
on the ratio $\sigma_L /\sigma_{T}$,
and it is further raised to the dotted curve due to the effect of (ii).
Comparing the dotted and the solid curve, we see that (iii) gives the 
enhancement, but it is not significant; the effects of (iii) 
affect $\sigma_{L}$ and $\sigma_{T}$ similarly,
so that they largely cancel for $\sigma_L /\sigma_{T}$.
Note the quite strong enhancement of the two-dot-dashed curve compared with the dot-dashed curve,
because of the significant overestimate of the suppression of $\sigma_{T}$ due to (i).

Another interesting aspect of 
Figs.~\ref{fig:L1s_Qdep_model}-\ref{fig:LT1s_Qdep_model}
is the characteristic behavior as a function of $Q^2$.
Rather different shape between $\sigma_{L}$ and $\sigma_{T}$ 
of Figs.~\ref{fig:L1s_Qdep_model} and \ref{fig:T1s_Qdep_model}
corresponds to the behavior of the ratio $\sigma_{L}/\sigma_{T}$ 
in Fig.~\ref{fig:LT1s_Qdep_model}, 
which increases almost linearly with increasing $Q^2$.
As is well-known, this $Q^2$-dependence is due to 
the behavior of the light-cone WF for the longitudinal photon, Eq.~(\ref{eqn:photon-wL}),
whose first term is proportional to $Q$ (see also Eq.~(\ref{eqn:7-1})).
Also, if we compare the explicit $\alpha$- or $\kvec$-dependent terms 
between Eqs.~(\ref{eqn:7-1}) and (\ref{eqn:7-2}), 
those terms reflecting (i), (ii)
depend differently on $\Qbar^2$. As we have shown above, however, 
the Fermi motion effects, produced via the convolution of Eq.~(\ref{eq:ampnew})
using Eqs.~(\ref{eqn:7-1}), (\ref{eqn:7-2}), 
do not significantly modify the shape of $\sigma_{L}$, $\sigma_{T}$, and $\sigma_{L}/\sigma_{T}$.
The Fermi motion effects mostly give overall enhancement or suppression,
by the different factors for $\sigma_{L}$ and $\sigma_{T}$ respectively.
In the next subsection, the ratio $\sigma_{L}/\sigma_{T}$, as well as
the corresponding total cross section $\sigma=\sigma_{L} + \sigma_{T}$,
will be compared with the recent HERA data.

\subsection{The $J/\psi$ photoproduction and electroproduction cross section}

Now, we present our numerical results for
the $J/\psi$ photo- and electroproduction cross section 
$\sigma(\gamma^{(*)}N\rightarrow J/\psi\ N)$ 
(see Eq.~(\ref{eqn:sigmatot})), including the new Fermi motion effects due to (i)-(iii),
and make a comparison with the available data.

In Figs.~\ref{fig:1s_Qdep} and \ref{fig:1s_Wdep},
we show $\sigma(\gamma^{*}N\rightarrow J/\psi\ N)$
as a function of $Q^{2}$ with $W=90$ GeV fixed,
and $\sigma(\gamma N\rightarrow J/\psi\ N)$
for the photoproduction ($Q^2=0$) as a function of $W$,
respectively.
In both figures, the dotted curve shows our result (\ref{eqn:sigmatot})
with Eqs.~(\ref{eq:ampnew}), (\ref{eqn:7-1}), (\ref{eqn:7-2}), and (\ref{eqn:scalar-wf}).
The solid curve is obtained by applying the replacement (\ref{eq:rescale})
to the dotted curve.
The dot-dashed and two-dot-dashed curves show the results 
without the effects due to (i)-(iii), using
the RRML vertex (\ref{eqn:IID-0}), (\ref{eqn:IID-1}) 
and the FKS vertex (\ref{eq:FKSL}), (\ref{eq:FKST})
for the effective $c\bar{c} \rightarrow V$ vertex, respectively.
Note that the dotted, dot-dashed, and two-dot-dashed curves
correspond to the cases of the solid, dot-dashed, and two-dot-dashed curves
in Figs.~\ref{fig:T1s_Qdep_model}, \ref{fig:LT1s_Qdep_model}, respectively.
We also show, by the dashed curve, the result obtained
from the dotted curve with the replacement (\ref{eq:static}),
in order to demonstrate the role of the Fermi motion effects
in the transverse direction.
%
\vspace*{0.5cm}
\begin{figure*}[htb]
\setlength{\unitlength}{1cm}
\begin{minipage}[t]{7.7cm}
\begin{picture}(6.5,6.5)
\hspace*{-0.5cm}
\psfig{file=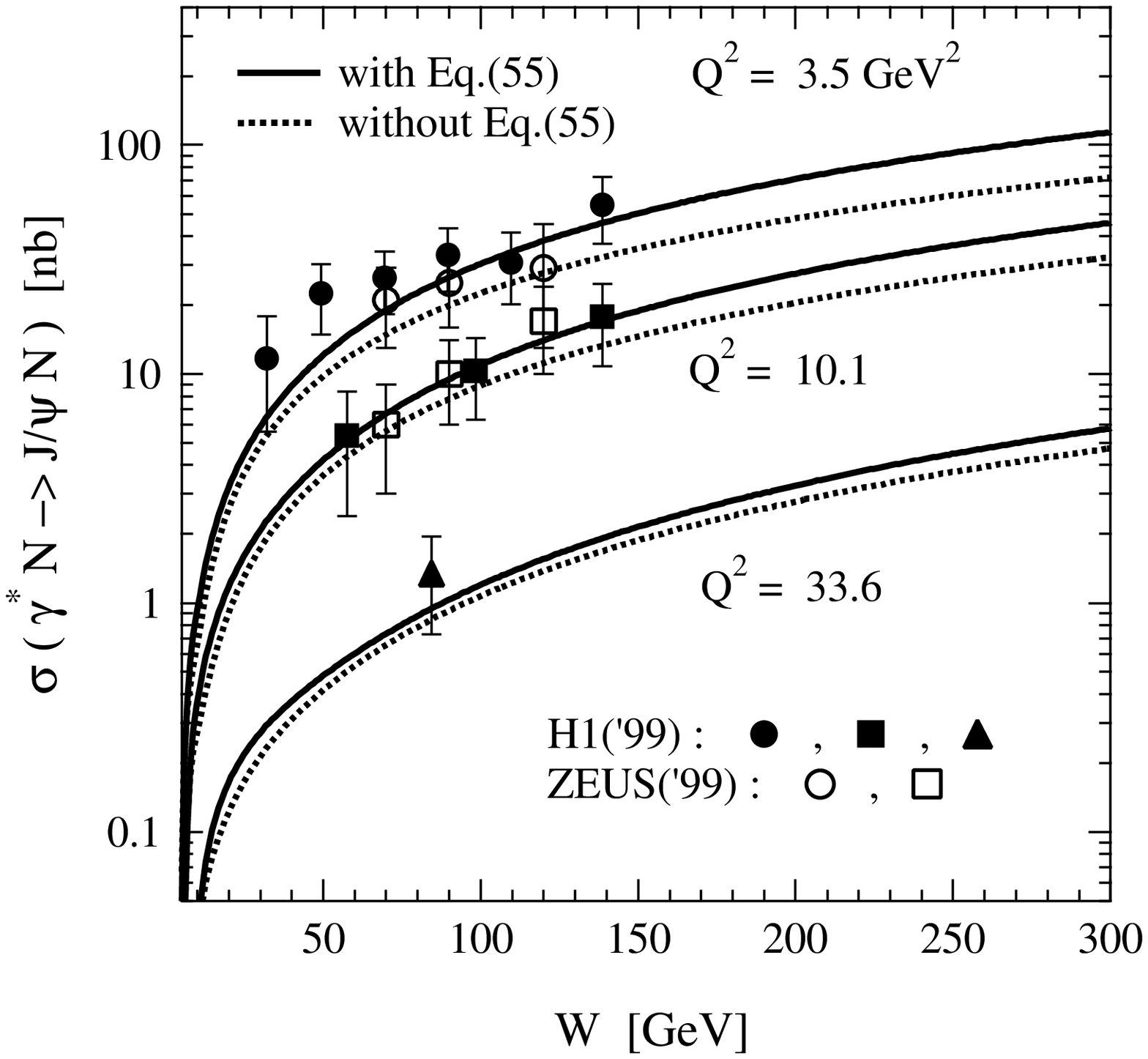,width=7.5cm,height=7.5cm}
\end{picture}\par
\caption{The total cross section for the electroproduction,
$\gamma^{*}N\rightarrow J/\psi\ N$, as a function of $W$ for
$Q^2= 3.5, 10.1$, and 33.6 GeV$^2$, in comparison with the 
recent HERA data \cite{H199,ZEUS99}.
The solid and the dotted lines have the same 
meaning as in Fig.~\ref{fig:1s_Wdep}.
The open squares show the ZEUS data at $Q^2=13$ GeV$^2$,
which is somewhat larger than the indicated value of $Q^{2}$.}
\label{fig:1s_Wdep_Qfnt}
\end{minipage}
\hspace*{0.7cm}
\begin{minipage}[t]{7.7cm}
\begin{picture}(6.5,6.5)
\hspace*{-0.5cm}
\psfig{file=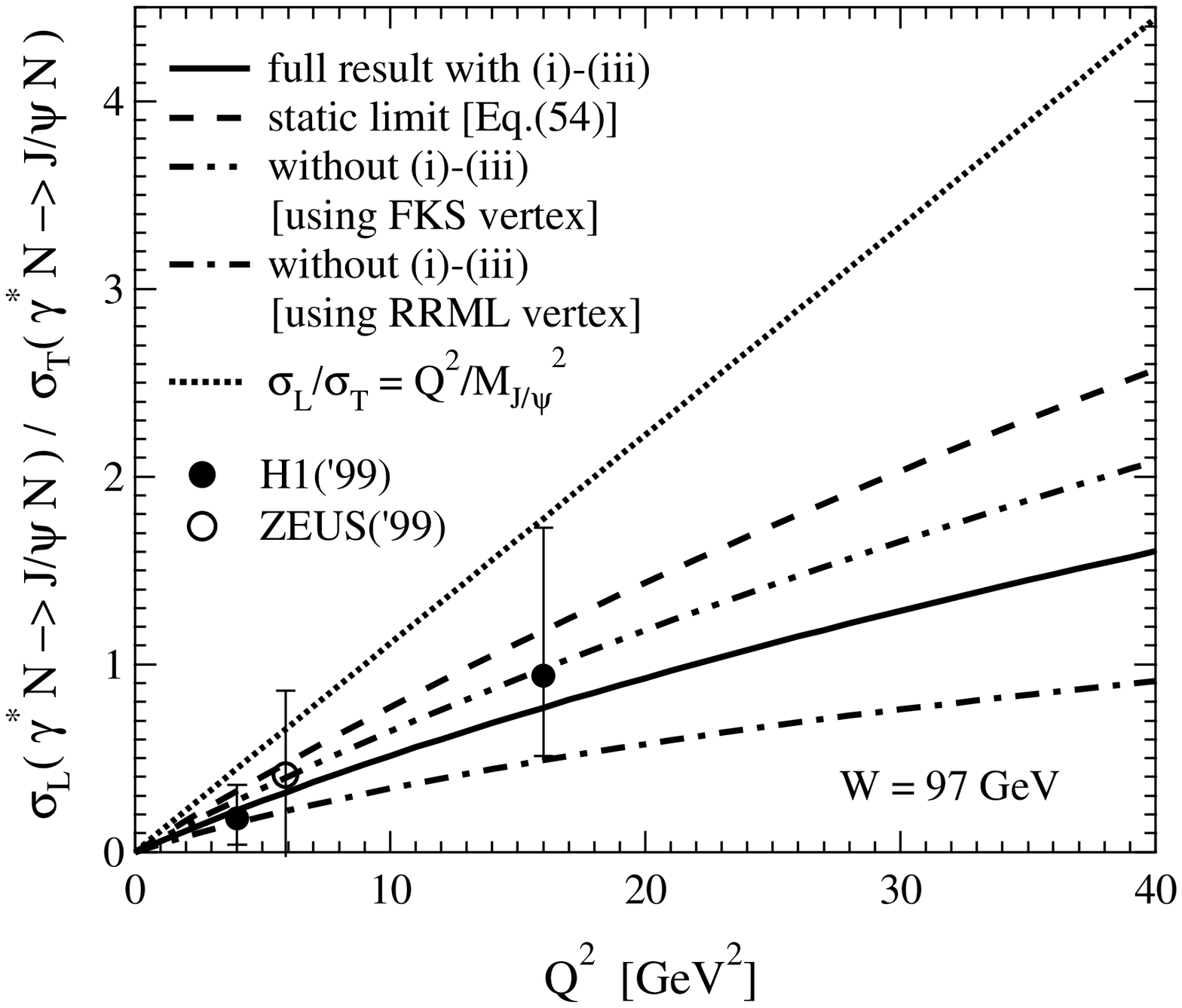,width=8cm,height=7.5cm}
\end{picture}\par
\caption{The ratio $\sigma_L/\sigma_T$ 
for the diffractive $J/\psi$ production in comparison with
the HERA data \cite{H199,ZEUS99}. 
The solid, dash-dotted, and two-dot-dashed curves are identical with the
corresponding curves in Fig.~\ref{fig:LT1s_Qdep_model}, respectively.
The dashed curve is obtained by applying the replacement (\ref{eq:static})
to the solid curve, while
the dotted line shows the result corresponding to Eq.~(\ref{eq:rys}).
}
\label{fig:LT1s_Qdep}
\end{minipage}
\end{figure*}

For $Q^{2} \lesssim 20$ GeV$^{2}$,
$\sigma_{T}$ is dominant in comparison with $\sigma_{L}$ 
(see Figs.~\ref{fig:L1s_Qdep_model}-\ref{fig:LT1s_Qdep_model}).
Therefore, the behavior of the curves of Fig.~\ref{fig:1s_Qdep} in this region
is very similar to that of Fig.~\ref{fig:T1s_Qdep_model}.
The behavior due to $\sigma_{L}$ shows up for $Q^{2} \gtrsim 20$ GeV$^{2}$, raising the tail. 
In particular, such characteristic behavior is clearly seen 
for the solid, dotted, and dot-dashed curves in Fig.~\ref{fig:1s_Qdep},
which are in agreement with the data. We note that the corresponding
three curves reproduce the behavior of the data also in Fig.~\ref{fig:1s_Wdep}.

One observes that the curves in 
Figs.~\ref{fig:1s_Qdep}, \ref{fig:1s_Wdep}
have some common features:

(a) For most region of $Q^2$ and $W$ shown in those figures, 
our full result (dotted curve) shows
slight ($10\sim 20$\%) suppression compared to the
dot-dashed curve, i.e., the case without (i)-(iii).
This reflects the corresponding suppression in Fig.~\ref{fig:T1s_Qdep_model},
because $\sigma \approx \sigma_{T}$ for $Q^{2} \lesssim 20$ GeV$^{2}$ in Fig.~\ref{fig:1s_Qdep}
and $\sigma = \sigma_{T}$ for Fig.~\ref{fig:1s_Wdep}. 
We note that the dot-dashed curve of Fig.~\ref{fig:1s_Wdep} coincides 
with the estimate of the Fermi motion effects for the $J/\psi$ photoproduction by RRML \cite{RRML},
up to the different form of the nonperturbative part of the light-cone WF, $\phi^{*}(\alpha, \kvec)$.

(b) The two-dot-dashed curve is suppressed very strongly in comparison with the other curves. 
This again reflects the corresponding strong suppression 
in Fig.~\ref{fig:T1s_Qdep_model},
due to the ``overestimate'' for the effect of (i) by using the FKS projector (\ref{eq:RFKS}).
We note that the two-dot-dashed curve in Figs.~\ref{fig:1s_Qdep}, \ref{fig:1s_Wdep}
coincides with the result calculated by FKS \cite{FKS},
up to some elaborate corrections considered in their paper,
such as the ``$Q^{2}$ rescaling'', the running quark mass, etc.

(c) The replacement (\ref{eq:rescale}) pushes up the dotted line to the solid line,
by a factor of $1.1\sim 1.5$ for Fig.~\ref{fig:1s_Qdep} 
and $1.2\sim 1.8$ for Fig.~\ref{fig:1s_Wdep}.
One may say that the ambiguity for the scale $Q^2_{\rm eff}$
would not lead to so significant modification of the results
for $W \lesssim 100$ GeV.
For higher $W$, the corresponding enhancement of the results
becomes pronounced, because the scale-dependence of the gluon
distribution $G(\tilde{x},Q^2_{\rm eff})$ becomes stronger
for smaller $\tilde{x}$ (see Eq.~(\ref{eq:tilx})).
We note that, if one applies the replacement (\ref{eq:rescale})
to the dot-dashed or the two-dot-dashed curves,
those curves are enhanced by almost the same factor
as the dotted curve. Thus, the two-dot-dashed curve
could be consistent with the data, too.
This might suggest that the NLO perturbative corrections
could give the effects of the same order as 
the Fermi motion corrections to the cross section for the charmonium production.

(d) Comparing the dotted and the dashed curves,
we recognize that the total contribution of the transverse Fermi motion gives
very strong suppression by a factor $\sim 1/4$.
This clearly shows the significant role of the transverse motion 
of the quarks for the charmonium production (see the discussion above Eq.~(\ref{eq:static})).

\vspace*{0.5cm}
\begin{figure*}[htb]
\setlength{\unitlength}{1.5cm}
\begin{minipage}[t]{7.7cm}
\begin{picture}(6.5,6.5)
\hspace*{-0.5cm}
\psfig{file=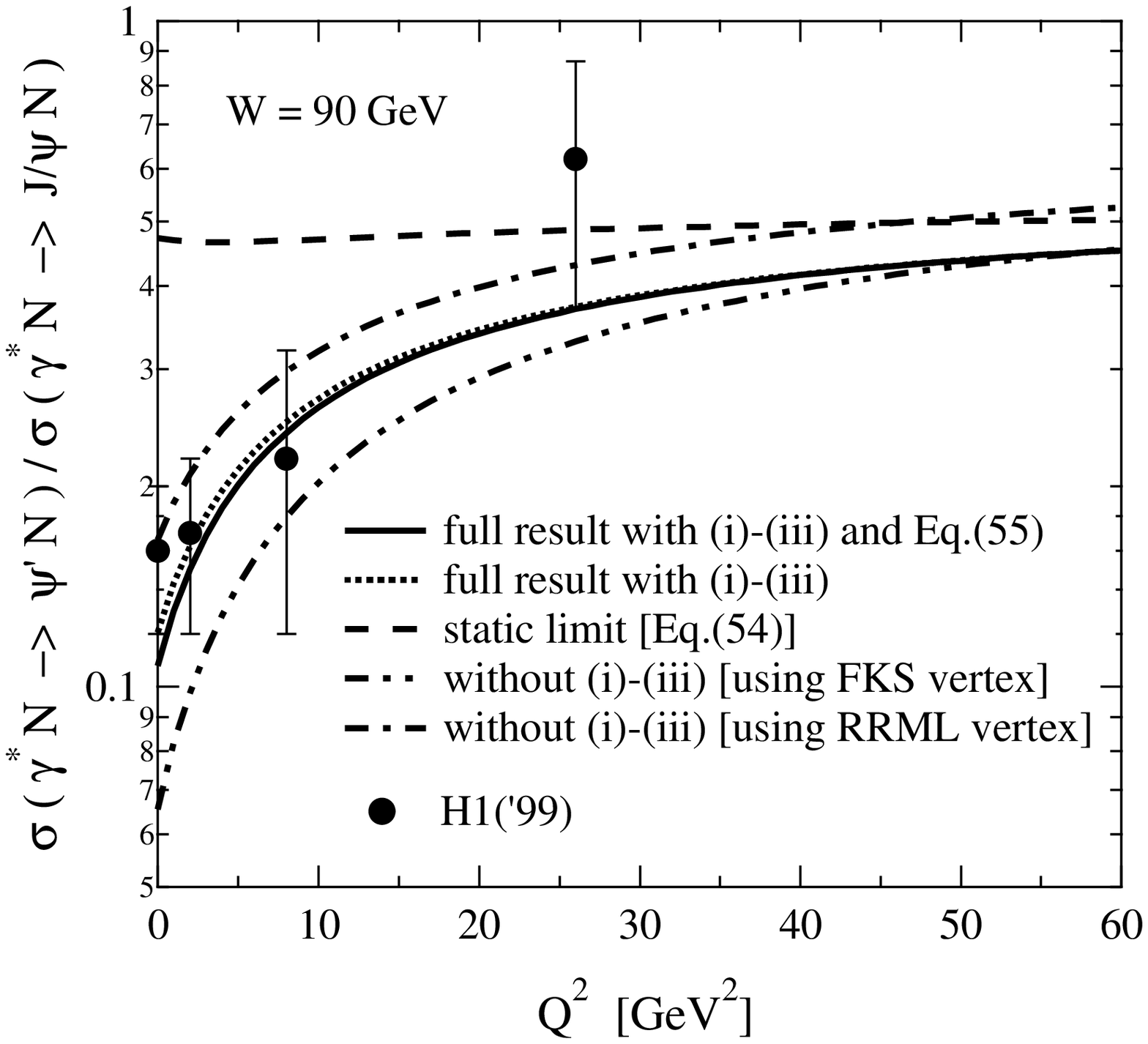,width=8cm,height=7.5cm}
\end{picture}\par
\caption{Same as Fig.~\ref{fig:1s_Qdep} but
for the ratio of the total cross section
for the electroproduction, $\gamma^{*}N\rightarrow \psi' N$,
to that for $\gamma^{*}N\rightarrow J/\psi\ N$.
}
\label{fig:2s1s_Qdep}
\end{minipage}
\hspace*{0.7cm}
\begin{minipage}[t]{7.7cm}
\begin{picture}(6.5,6.5)
\hspace*{-0.5cm}
\psfig{file=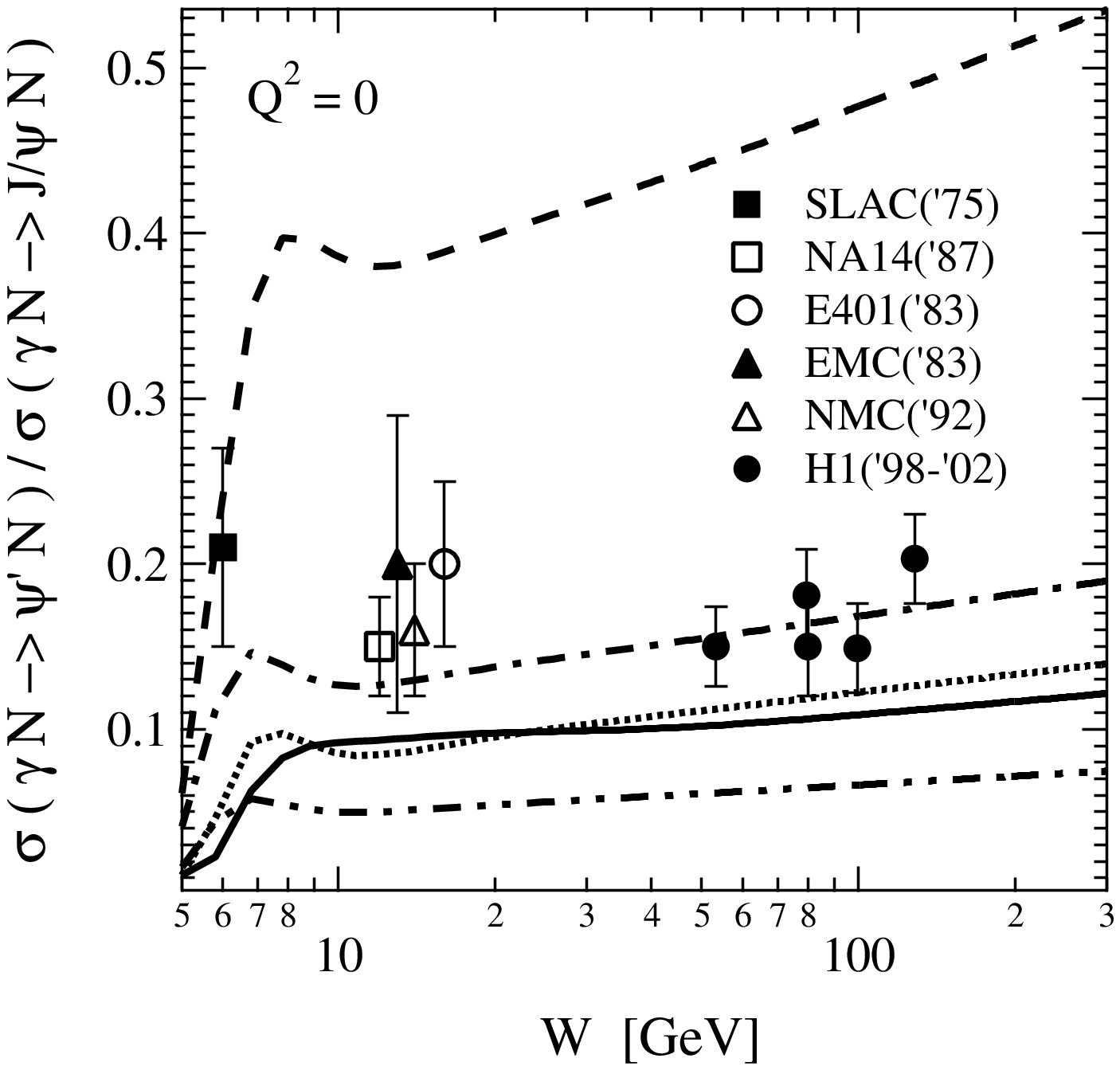,width=8cm,height=7.5cm}
\end{picture}\par
\caption{Same as Fig.~\ref{fig:1s_Wdep} but
for the ratio of the total cross section
for the photoproduction, $\gamma N\rightarrow \psi' N$,
to that for $\gamma N\rightarrow J/\psi\ N$.
Experimental data points from
earlier fixed target \cite{SLAC,NA14,E401b,EMC,NMC} and the H1 \cite{H198,H102} experiments. 
}
\label{fig:2s1s_Wdep}
\end{minipage}
\end{figure*}

%
Fig.~\ref{fig:1s_Wdep_Qfnt} shows the $J/\psi$ electroproduction cross section, 
$\sigma(\gamma^{*} N\rightarrow J/\psi\ N)$, as a function of $W$
for $Q^2=3.5$, $10.1$, and $33.6$ GeV$^2$. 
The solid and the dotted lines have the same 
meaning as in Fig.~\ref{fig:1s_Wdep}.
The $W$ dependence of the results is similar 
to the photoproduction case in Fig.~\ref{fig:1s_Wdep}, 
and both solid and dotted curves reproduce the data.
Although we have not shown explicitly,
we have also calculated the results corresponding to the dot-dashed as well as the
two-dot-dashed curve of Fig.~\ref{fig:1s_Wdep}, and find that
those results are also consistent with the data in Fig.~\ref{fig:1s_Wdep_Qfnt}.

Fig.~\ref{fig:LT1s_Qdep} compares 
the results of $\sigma_L/\sigma_T$ of Fig.~\ref{fig:LT1s_Qdep_model}
with the recent HERA data for $W=97$ GeV \cite{H199,ZEUS99}.
The solid, dash-dotted, and two-dot-dashed curves are identical with the
corresponding curves in Fig.~\ref{fig:LT1s_Qdep_model}, respectively.
The dashed curve is obtained by applying the replacement (\ref{eq:static})
to the solid curve, while
the dotted line shows the result corresponding to Eq.~(\ref{eq:rys}),
i.e., $\sigma_L/\sigma_T$$=Q^2/M_{J/\psi}^2$. 
Thus, the total contribution due
to the Fermi motion is significant, and the quantitative role of the transverse quark motion is similar to that of the longitudinal one.

The results with the replacement (\ref{eq:rescale})
are not shown in Fig.~\ref{fig:LT1s_Qdep}.
We actually find that, for the ratio $\sigma_L/\sigma_T$,
the effects due to the replacement (\ref{eq:rescale})
almost cancel between the numerator and the denominator.
This suggests that the ratio $\sigma_L/\sigma_T$ would 
allow  a ``clean'' theoretical prediction insensitive to the ambiguity for the scale $Q_{\rm eff}^{2}$.
One might further expect that $\sigma_L/\sigma_T$ could be insensitive
even to other ambiguities or corrections like the uncertainties 
in the $\tilde{x}$-dependence of $G(\tilde{x}, Q_{\rm eff}^{2})$ 
and the NLO perturbative corrections to the diffractive amplitude.
Moreover, as demonstrated in Sec.~III A,
the strong helicity-dependence of the new Fermi motion effects due to (i), (ii)
modifies $\sigma_L/\sigma_T$ strikingly, so that we expect that the ratio $\sigma_L/\sigma_T$ could be the most suitable discriminator among the predictions 
of the Fermi motion effects by various models.
Unfortunately, however, the available data in Fig.~\ref{fig:LT1s_Qdep} 
are not enough for this purpose.

\subsection{The $\psi'$ photoproduction and electroproduction cross section}

It is straightforward to extend our calculation in Sec.~III B to the $\psi'$ production, by substituting the solution for the first radial excitation 
of the relevant Schr\"{o}dinger equation 
into $\phi_{NR}$ of Eq.~(\ref{eqn:scalar-wf}), 
and by the trivial replacements, $M_{J/\psi}\rightarrow M_{\psi'}$,
$B_{J/\psi}\rightarrow B_{\psi'}$, in the relevant formulae.
We discuss our numerical results for
the $\psi'$ photo- and electroproducion cross section
$\sigma(\gamma^{(*)}N\rightarrow \psi' N)$, including the new Fermi motion effects 
due to (i)-(iii), and also make a comparison with the available data
on the ratio of $\psi'$ to $J/\psi$ cross section.

In Fig.~\ref{fig:2s1s_Qdep},
we show the ratio 
$\sigma(\gamma^{*}N\rightarrow \psi' N)/\sigma(\gamma^{*}N\rightarrow J/\psi\ N)$
for the electroproduction as a function of $Q^{2}$ at $W=90$ GeV.
The lines have the same meaning as in Fig.~\ref{fig:1s_Qdep}: 
the solid curve shows the ratio of $\sigma(\gamma^{*}N\rightarrow \psi' N)$, 
which is calculated similarly to the solid curve in Fig.~\ref{fig:1s_Qdep}, 
to $\sigma(\gamma^{*}N\rightarrow J/\psi\ N)$,
which is given by the solid curve of Fig.~\ref{fig:1s_Qdep};
the dotted curve shows the ratio of $\sigma(\gamma^{*}N\rightarrow \psi' N)$, which is calculated similarly to the dotted curve in Fig.~\ref{fig:1s_Qdep}, 
to $\sigma(\gamma^{*}N\rightarrow J/\psi\ N)$,
which is given by the dotted curve of Fig.~\ref{fig:1s_Qdep};
and so on.

We see that the solid curve almost coincides with the dotted curve,
because, similarly to the ratio 
$\sigma_{L}/\sigma_{T}$ of Fig.~\ref{fig:LT1s_Qdep},
the effects due to the replacement (\ref{eq:rescale})
almost cancel between the numerator and the denominator.
Note that, for the dotted curve of the present case,
we use $Q_{\rm eff}^2 = (Q^{2}+M_{J/\psi}^{2})/4$ 
for $\alpha_{s}(Q_{\rm eff}^2)G(\tilde{x}, Q_{\rm eff}^2)$ 
of the denominator,
and $Q_{\rm eff}^2 = (Q^{2}+M_{\psi'}^{2})/4$ 
for that of the numerator (see Eqs.~(\ref{eqn:7-1}), (\ref{eqn:7-2}));
actually, this difference of $Q_{\rm eff}^{2}$ 
between $J/\psi$ and $\psi'$
gives the small difference of the solid and the dotted curves for the low $Q^{2}$ region.
Apparently, such small difference is irrelevant, and 
for high $Q^{2}$, i.e., $Q^2 \gg M_{J/\psi}^2, M_{\psi'}^2$, the difference between those two curves disappears.
However, the corresponding difference of $Q_{\rm eff}^{2}$ could
produce relevant behavior for other observables
(see the discussion about Fig.~\ref{fig:2s1s_Wdep} below).

%
\begin{figure*}[htb]
\setlength{\unitlength}{1.5cm}
\begin{minipage}[t]{7.7cm}
\begin{picture}(6.5,6.5)
\hspace*{-0.5cm}
\psfig{file=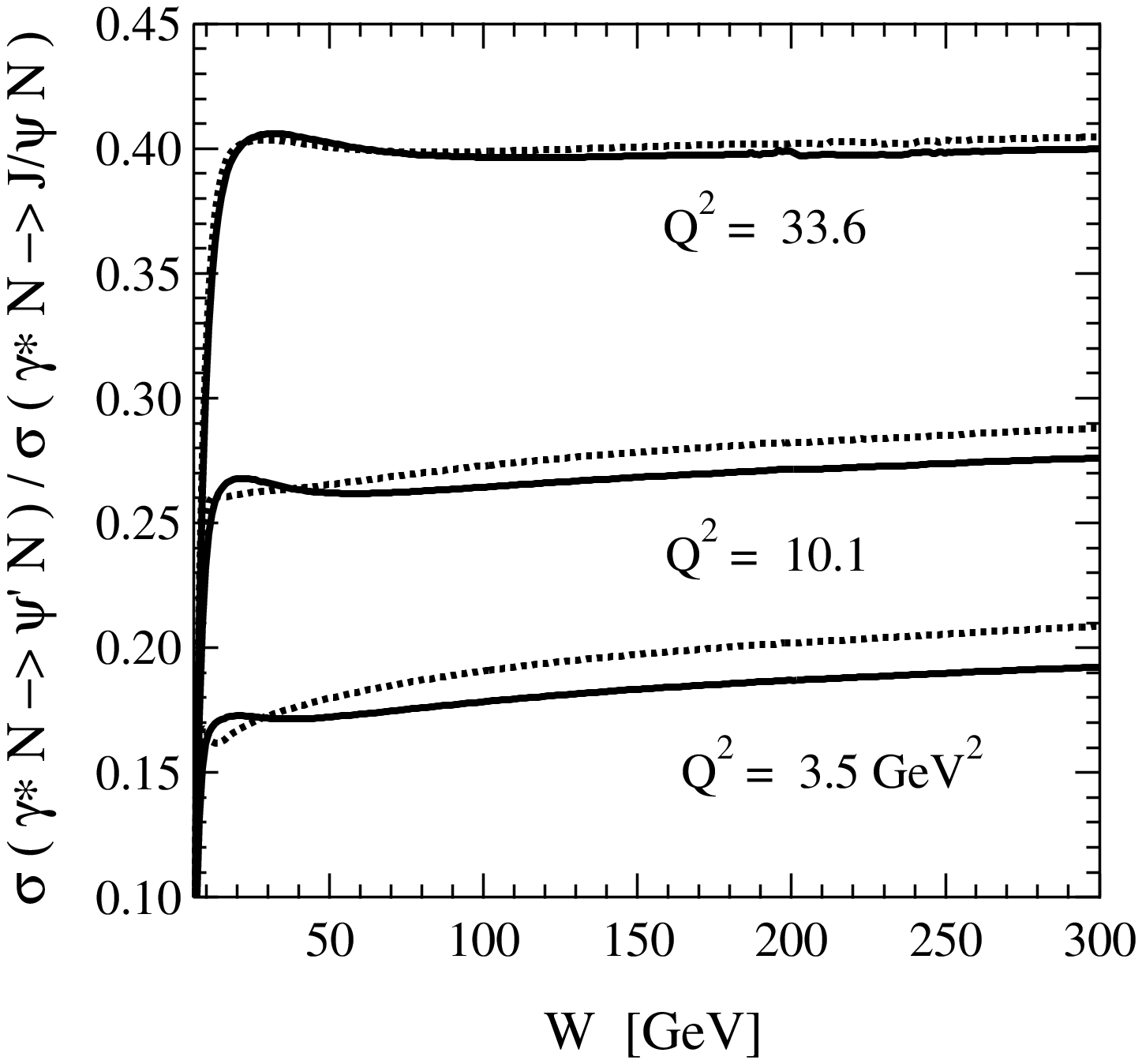,width=7.5cm,height=7.5cm}
\end{picture}\par
\caption{Same as Fig.~\ref{fig:1s_Wdep_Qfnt} but
for the ratio of the total cross section
for the electroproduction, $\gamma^{*}N\rightarrow \psi' N$,
to that for $\gamma^{*}N\rightarrow J/\psi\ N$.
}
\label{fig:2s1s_Wdep_Qfnt}
\end{minipage}
\hspace*{0.7cm}
\begin{minipage}[t]{7.7cm}
\begin{picture}(6.5,6.5)
\hspace*{-0.5cm}
\psfig{file=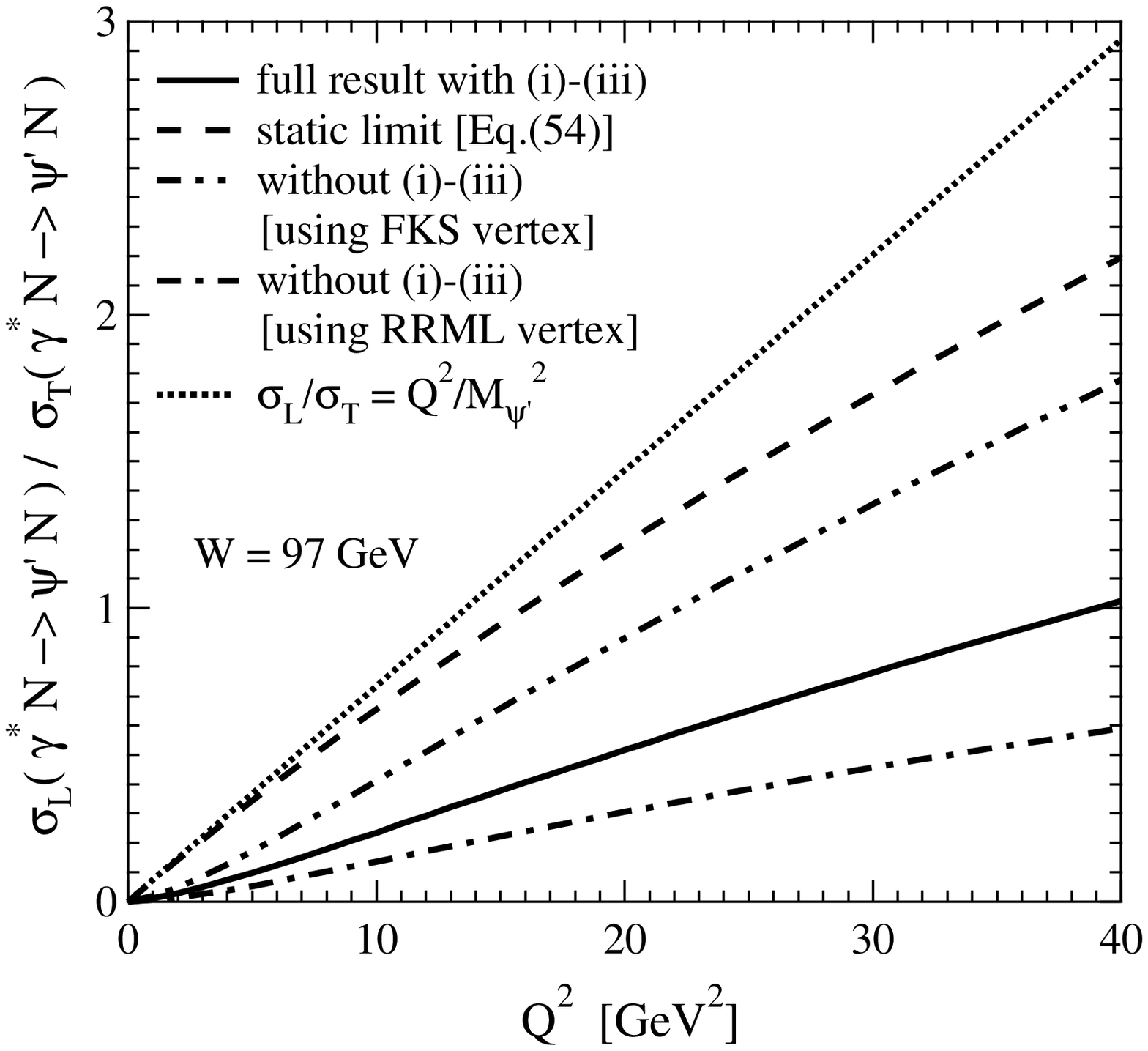,width=8cm,height=7.5cm}
\end{picture}\par
\caption{Same as Fig.~\ref{fig:LT1s_Qdep} 
but for the diffractive $\psi'$ production.}
\label{fig:LT2s_Qdep}
\end{minipage}
\end{figure*}

The comparison of the dashed line with the dotted line
suggests that the Fermi motion effects are more pronounced
for the $\psi'$ production than for the $J/\psi$ one.
In fact, the typical velocity $v$ of Eq.~(\ref{eq:v2})
is larger for the ``2$S$-state'' $\psi'$ than for the ``1$S$-state'' $J/\psi$.
We also recognize an interesting point:
only the dashed line has the ``flat'' behavior,
while all the other lines have qualitatively a similar shape with each other,
decreasing steeply with decreasing $Q^{2}$.
It is easy to see that this latter behavior is characteristic 
of the Fermi motion effects for the case of the radial excitation, 
reflecting that the corresponding meson WF has a ``node'':
for $\sigma(\gamma^{*}N\rightarrow \psi' N)$,
we calculate the convolution (\ref{eq:ampnew}) 
with the nonperturbative WF $\phi^{*}(\alpha, \kvec)$ 
for $\psi'$, which is given as  
Eq.~(\ref{eqn:scalar-wf}) in terms of the nonrelativistic, 
2$S$-state WF $\phi_{NR}$.
Because $\phi_{NR}$ has the node at $r\sim 1/(m_{c} v)$ in the coordinate space, 
the convolution (\ref{eq:ampnew}) suffers from the strong cancellation.
We note that the behavior of $\Omega^{(\zeta)}(\alpha, \kvec)$ as a function of $\alpha, \kvec$ is controlled by the scale $Q^2$ (see Eqs.~(\ref{eqn:7-1}), (\ref{eqn:7-2}));
therefore, the cancellation can be avoided when $Q^{2} \gg m_{c}^{2}$, 
i.e., $\Omega^{(\zeta)}$ is ``localized'' sufficiently in the coordinate space representation.
This effect leads to the steep $Q^2$-dependence 
for $\sigma(\gamma^{*}N\rightarrow \psi' N)$,
which is absent for $\sigma(\gamma^{*}N\rightarrow J/\psi\ N)$, so that one 
obtains the behavior of the curves in Fig.~\ref{fig:2s1s_Qdep}.
The reason why the dashed curve has the flat behavior
is that we use the replacement (\ref{eq:static}) for the $\psi'$ WF in the numerator,
as well as for the $J/\psi$ WF in the denominator, to calculate 
$\sigma(\gamma^{*}N\rightarrow \psi' N)/\sigma(\gamma^{*}N\rightarrow J/\psi\ N)$
in this case.

Our full result (dotted or solid line), including the new Fermi motion effects due to (i)-(iii),
is suppressed by a factor of about $2/3$ than the dot-dashed line,
which corresponds to the case without (i)-(iii)
and uses the RRML vertex (\ref{eqn:IID-0}), (\ref{eqn:IID-1}).
On the other hand, the two-dot-dashed line, which corresponds to another case without (i)-(iii),
shows much stronger suppression.
This is due to the combined effect of the ``overestimate'' of 
the effect of (i) by using the FKS projector (\ref{eq:RFKS}),
and of the fact that the Fermi motion effects are more pronounced
for the radial excitation $\psi'$ than for $J/\psi$.

In Fig.~\ref{fig:2s1s_Wdep}, we show the ratio 
$\sigma(\gamma N\rightarrow \psi' N)/\sigma(\gamma N\rightarrow J/\psi\ N)$
for the photoproduction ($Q^2=0$) as a function of $W$.
The lines have the same meaning as in Fig.~\ref{fig:2s1s_Qdep}.  
For almost the whole region of $W$, we see the behavior similar to that observed for $Q^{2} \rightarrow 0$ in Fig.~\ref{fig:2s1s_Qdep}.
In particular, the dashed curve with the replacement (\ref{eq:static})
largely overestimates the data by a factor of $2\sim 3$, 
in contrast with the other curves.
This suggests that the strong transverse Fermi motion effects, reflecting
the radial shape of the 2$S$-state WF, is essential 
to explain the tendency of the data.
On the other hand, the two-dot-dashed curve, using the FKS vertex (\ref{eq:FKSL}), (\ref{eq:FKST}),
underestimates the data by a factor of $1/3\sim 1/2$. 
We will add some comments on this point
in the final part of this subsection.

Our full result, as well as all the other curves in Fig.~\ref{fig:2s1s_Wdep}, 
shows a slight increase with increasing $W$, above $W\simeq 10$ GeV, and
this behavior seems to match with the tendency of the recent H1 data \cite{H102}.
In the present calculation,
this weak $W$ dependence originates from
the difference of $Q_{\rm eff}^{2}= (Q^{2} + M_{V}^{2})/4$
in the gluon distribution $G(\tilde{x}, Q_{\rm eff}^2)$,
corresponding to $V=\psi'$ and $J/\psi$:
the larger $Q_{\rm eff}^{2}$ gives the faster increase of 
$G(\tilde{x}, Q_{\rm eff}^2)$ 
with decreasing $\tilde{x}$ (see Eq.~(\ref{eq:tilx})).
Actually, this fact motivates our choice as $Q_{\rm eff}^{2}= (Q^{2} + M_{V}^{2})/4$
following Ref.~\cite{Ryskin},
although, within the LLA, other choice is possible.
For example, in Ref.~\cite{IIvanov}, 
$Q_{\rm eff}^{2}= (Q^{2} + 4m_{c}^{2})/4$ was used.
We have checked that this choice makes the $W$ dependence of the ratio 
$\sigma(\gamma N\rightarrow \psi' N)/\sigma(\gamma N\rightarrow J/\psi\ N)$
almost constant. 
Another choice used in the literature 
is $Q_{\rm eff}^2=\alpha(1-\alpha)Q^2+m_c^2+\kvec^2$
\cite{RRML}. In this case, we find that 
the ratio 
$\sigma(\gamma N\rightarrow \psi' N)/\sigma(\gamma N\rightarrow J/\psi\ N)$ 
shows a weak rise 
similarly to that 
in Fig.~\ref{fig:2s1s_Wdep}, due to the convolution with the $\psi'$ or $J/\psi$ WF 
$\phi^{*} (\alpha, \kvec)$ with respect to $\kvec$.

A similar rise as a function of $W$ is observed in Fig.~\ref{fig:2s1s_Wdep_Qfnt}, which shows the ratio 
$\sigma(\gamma^{*}N\rightarrow \psi' N)/\sigma(\gamma^{*}N\rightarrow J/\psi\ N)$
for the electroproduction for $Q^{2}=3.5, 10.1$, and 33.6 GeV$^{2}$.
The solid and the dotted lines have the same meaning as those in Fig.~\ref{fig:2s1s_Wdep}.

Now, we go over to the helicity dependence in the diffractive $\psi'$ production.
Fig.~\ref{fig:LT2s_Qdep} shows the ratio
$\sigma_{L}(\gamma^{*}N\rightarrow \psi' N)
/\sigma_{T}(\gamma^{*}N\rightarrow \psi' N)$
for the electroproduction of $\psi'$ as a function of $Q^{2}$ at $W=97$ GeV.
The lines have the same meaning as in Fig.~\ref{fig:LT1s_Qdep} 
for the $J/\psi$ production.
We see that the strong helicity-dependence observed in Fig.~\ref{fig:LT1s_Qdep}
is even more pronounced for the $\psi'$ production.
We expect that such behavior is insensitive to elaborate corrections,
and $\sigma_{L}/\sigma_{T}$ for the $\psi'$ production allows a clean prediction,
although we do not have the experimental data to be compared.

\begin{figure}[h]
\vspace*{0cm}
\begin{center}
\hspace*{0cm}
\psfig{file=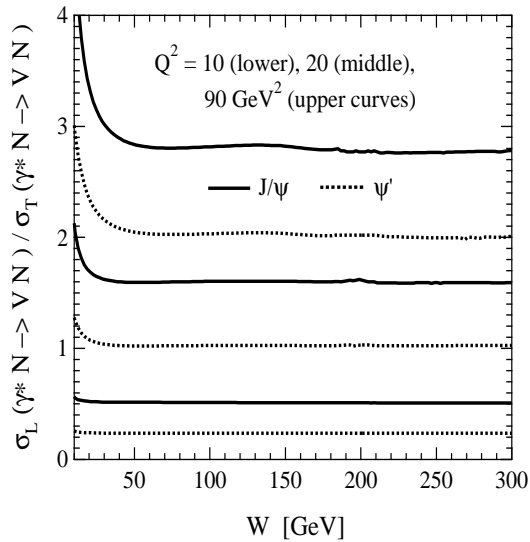,width=7.5cm,height=7.5cm}
\end{center}
\vspace{0cm}
\caption{The ratio of the cross sections for the diffractive vector-meson production
with longitudinally and transversely polarized photons, 
as a function of $W$ at $Q^2= 10, 20$, and 90 GeV$^2$.
The solid lines show the $J/\psi$ production, while the dotted lines
show the $\psi'$ production.
All lines are calculated using Eqs.~(\ref{eqn:sigmaL}), (\ref{eqn:sigmaT})
with the replacement (\ref{eq:rescale}).
}
\label{fig:LT1s_2s_Wdep_Qfnt}
\end{figure}

%
%
We also show, in Fig.~\ref{fig:LT1s_2s_Wdep_Qfnt},
the $W$ dependence of the ratio 
$\sigma_{L}(\gamma^{*}N\rightarrow \psi' N)
/\sigma_{T}(\gamma^{*}N\rightarrow \psi' N)$
by the dotted lines, and of 
$\sigma_{L}(\gamma^{*}N\rightarrow J/\psi\ N)
/\sigma_{T}(\gamma^{*}N\rightarrow J/\psi\ N)$
by the solid lines,
for $Q^{2}=10, 20$, and 90 GeV$^{2}$.
All lines are calculated using our full results (\ref{eqn:sigmaL}), (\ref{eqn:sigmaT})
with the replacement (\ref{eq:rescale}).
We see that those results are almost independent of $W$
for the energy region available at HERA.
This implies that the relevant $W$ dependence through $\tilde{x}$ 
of Eqs.~(\ref{eqn:7-1}), (\ref{eqn:7-2})
cancels with each other between 
the denominator and the numerator of $\sigma_{L}/\sigma_{T}$,
even though we take into account the $\alpha$ and $\kvec$ dependence of $\tilde{x}$ 
as in Eq.~(\ref{eq:tilx}).
When we make the replacement (\ref{eq:x}),
the approximately constant behavior of the $\sigma_L/\sigma_T$ ratio 
as a function of $W$ would be an immediate consequence of the cancellation 
of the gluon distribution $G(x, Q_{\rm eff}^{2})$
between the denominator and the numerator, 
as is also seen from the corresponding result of Ref.~\cite{MRT}.

Finally, we mention some recent
works which also discussed the ratio of $\psi'$ to $J/\psi$ cross sections.
In Figs.~\ref{fig:2s1s_Qdep}, \ref{fig:2s1s_Wdep},
we have shown the strong suppression of the two-dot-dashed curve 
compared with the data.
Such strong suppression  
with the use of the FKS vertex (\ref{eq:FKSL}), (\ref{eq:FKST})
was first claimed for the photoproductions in Ref.~\cite{HP}.
In the present work,
we have clarified that the main reason of this too strong suppression 
is the ``overestimate'' of 
the Fermi motion effect of (i) 
by using the ``oversimplified'' FKS projector (\ref{eq:RFKS}).
So far, there exist some works \cite{HIKT,SHIAH}, which suggested 
other mechanism to avoid the corresponding strong suppression.
Ref.~\cite{HIKT} considered 
the role of the ``new'' spin structure for the $^{3}S_{1}$ $c\bar{c}$ state 
originating from the Melosh spin rotation,
which is to relate the quantities in the constituent quark model 
to those in the infinite momentum frame.
Ref.~\cite{SHIAH} discussed the role of the higher order terms 
in the power series of $\lvec\cdot \hat{\bvec}$ in the integrand of Eq.~(\ref{eqn:A}), 
which corresponds to going beyond the LLA, i.e., the color-dipole picture.
The detailed comparison of our results 
with those of Refs.~\cite{HIKT,SHIAH} is beyond the scope of this work.

\section{Summary and discussion}
In the present paper,
we have reexamined the Fermi motion corrections 
to the diffractive photo- and electroproductions 
of the heavy vector-mesons, $J/\psi$ and $\psi'$, in the LLA of pQCD. 
We have taken into account all the Fermi motion corrections
arising from the relative motion of quarks inside
the charmonium, which is treated as a nonrelativistic bound state
of $c$ and $\bar{c}$.
The key ingredients in our approach are
the projector ${\cal R}$
to ensure the spin structure for the $^{3}S_{1}$ $c\bar{c}$ bound state,
the off-shellness in the hadronization ($c\bar{c} \rightarrow V$) vertex,
and the modification of the Bjorken-$x$, i.e., 
the gluon's longitudinal momentum fraction probed by the process,
due to the coupling with the $c\bar{c}$ pair in internal relative motion.
Although these three contributions were not considered properly in the previous works,
we have demonstrated that all these contributions
should be included in the LLA diffractive amplitude in QCD, 
and that all of them produce the new Fermi motion effects
of order $v^2$ with $v$ being the heavy-quark velocity inside the vector meson.

We have presented a quantitative estimate of the Fermi motion effects
on the diffractive production cross sections, including our three new contributions.
In the QCD factorization formula for the diffractive charmonium production,
the nonperturbative, internal motion of quarks 
is represented by the light-cone WFs for the charmonium,
which we have constructed from the corresponding $S$-wave solution of 
the Schr\"odinger equation with a realistic potential 
between $c$ and $\bar{c}$. 
We find that the net effect from the Fermi motion gives moderate suppression 
of the total cross section for the $J/\psi$ photo- and electroproductions.
The corresponding suppression is similar to that obtained by RRML \cite{RRML} for the 
$J/\psi$ photoproduction,
but it is much weaker than that observed by FKS \cite{FKS}
for the photoproduction as well as the electroproduction.
We have also clarified the strong helicity-dependence of our new Fermi motion effects,
and predicted novel behavior in the helicity-dependent cross sections:
the slope of the longitudinal to transverse production ratio $\sigma_L/\sigma_T$
as a function of $Q^{2}$ distinguishes clearly our result from 
the results corresponding to the previous works.
Because the $\sigma_L/\sigma_T$ ratio is expected to be insensitive to
elaborate corrections or theoretical uncertainties, we propose 
it as a clean discriminator among the various predictions of
the Fermi motion effects. 
For the detailed comparison of the 
$\sigma_L/\sigma_T$ ratio between theory and experiment,
we should wait for the new precise data, especially for the high $Q^2$ region.

We have also observed the similar behavior for the $\psi'$ photo- and electroproductions.
Our results for the ratio of $\psi'$ to $J/\psi$ cross sections 
indicate that the suppression of the cross section becomes somewhat stronger for the $\psi'$ 
production than for the $J/\psi$ one, reflecting rapider motion of quarks
inside the first radial excitation.
We have demonstrated that 
the $\psi'$ to $J/\psi$ ratio is also useful 
in discriminating the various theoretical calculations
of the Fermi motion effects.
Another interesting point is that 
the $Q^{2}$ dependence of the $\psi'$ to $J/\psi$ ratio
is sensitive to the detailed shape 
of the WFs of the $\psi'$ as well as of the $J/\psi$,
while it is insensitive to the behavior of the gluon distribution.

As a novel feature in the heavy meson production,
we have emphasized that
the Fermi motion in the transverse direction is as important as that
in the longitudinal direction.
The transverse quark motion plays a major role to
produce characteristic effects due to the Fermi motion,
especially the strongly helicity-dependent effects.

The comparison of our results with the HERA data has been made in this paper.
Taking into account the theoretical ambiguity associated with the LLA, 
we may conclude that our full results involving the new Fermi motion effects 
are in agreement with the recent data.
However,
it should be taken with reservation, because, in this work, we have
not pursued some corrections, which could in principle
produce the effects of the same order as the Fermi motion corrections.
Those include genuine relativistic corrections due to
the ``small components'' of the meson WF $\Psi^{V}$,
and the next-to-leading perturbative corrections to the photon WF $\Psi^{\gamma}$
and the $c\bar{c}$-gluon hard scattering amplitude ${\cal A}^{c\bar{c}g}$ 
(see Eqs. (\ref{eqn:a}), (\ref{eqn:a2})).
Leaving aside those unresolved problems for the present,
the comparison has shown that our results reproduce the overall behavior of the data
as functions of $Q^2$ as well as the energy $W$,
over a range of observables.
Therefore, what we can learn from this study is that 
the Fermi motion effects provide natural mechanisms within the LLA,
which give the characteristic behavior of
the total cross sections as well as of the helicity-dependent cross sections
observed in experiment.
For the more detailed quantitative comparison with the data, 
one would eventually have to 
take into account further sophisticated corrections besides above-mentioned ones:
those include, e.g., the contribution of higher 
Fock states such as $|c\bar{c}g\rangle$ 
in the charmonia,
the ``skewedness'' 
of the gluon distribution \cite{MR},
and the $W$-dependence of the $t$-slope parameter $B_V$ 
for the differential cross section $d\sigma/dt$
\cite{FMS}.

\acknowledgements
We are grateful to T.~Hatsuda for useful discussions 
at the early stage of this work, 
and to K.~Suzuki for helpful comments. 
The work of A.H. was supported by Alexander von Humboldt Research Fellowship.
The work of K.T. was supported in part by the Grant-in-Aid of the Sumitomo Foundation.


\end{document}